\documentclass[11pt, a4paper]{article}
\usepackage{jcappub}

\usepackage{aasmacros}
\usepackage{booktabs}
\usepackage{multirow}
\usepackage{amssymb}
\usepackage{floatrow}
\newfloatcommand{capbtabbox}{table}[][3.5in]
\newfloatcommand{capbfigbox}{figure}[][2.3in]
\usepackage{blindtext}
\usepackage{amsmath}
\usepackage[normalem]{ulem}
\usepackage{hyperref}
\usepackage{soul}

\newcommand{\Fermi}{\textit{Fermi}}

\newcommand{\Msun}{$M_\odot$}
\newcommand{\sv}{\langle \sigma v \rangle}
\newcommand{\Jf}{$\mathcal{J}$-factor}

\def \aap  {A\&A}

\def \apjs  {ApJS}
\def \apjl  {ApJL}

\begin{document}

\begin{flushright}
LAPTH-064/16\\
IFT-UAM/CSIC-16-118\\
FTUAM-16-40\\
\end{flushright}
\title{Realistic estimation for the detectability of dark matter sub-halos with {\it Fermi}-LAT}

\author[a]{Francesca Calore}
\emailAdd{francesca.calore@lapth.cnrs.fr}
\author[b]{Valentina De Romeri}
\emailAdd{valentina.deromeri@uam.es}
\author[c]{Mattia Di Mauro}
\emailAdd{mdimauro@slac.stanford.edu}
\author[d,e]{Fiorenza Donato}
\emailAdd{donato@to.infn.it}
\author[f]{Federico Marinacci}
\emailAdd{fmarinac@mit.edu}

\affiliation[a]{LAPTh, CNRS, 9 Chemin de Bellevue, B.P.110 Annecy-le-Vieux, F-74941, France}
\affiliation[b]{Departamento de F\'{\i}sica Te\'orica and Instituto de F\'{\i}sica Te\'orica, IFT-UAM/CSIC,\\
Universidad Aut\'onoma de Madrid, Cantoblanco, 28049 Madrid, Spain}
\affiliation[c]{W. W. Hansen Experimental Physics Laboratory, Kavli Institute for Particle Astrophysics and Cosmology, 
Department of Physics and SLAC National Accelerator Laboratory, \\ Stanford University, Stanford, CA 94305, USA}
\affiliation[d]{Dipartimento di Fisica, Universit\`a di Torino, via P. Giuria 1, I-10125 Torino, Italy}
\affiliation[e]{Istituto Nazionale di Fisica Nucleare, Sezione di Torino, via P. Giuria 1, I-10125 Torino, Italy}
\affiliation[f]{Kavli Institute for Astrophysics and Space Research, Massachusetts Institute of Technology, Cambridge, MA 02139, USA}

\date{\today}

\abstract{
Numerical simulations of structure formation have recorded a remarkable progress in the recent years, 
in particular due to the inclusion of baryonic physics evolving with the dark matter component. 
We generate Monte Carlo realizations of the dark matter sub-halo population 
based on the results of the recent hydrodynamical simulation suite of Milky Way-sized galaxies~\cite{Marinacci:2013mha}.
We then simulate the gamma-ray sky for both the setup of the 3FGL and 2FHL \Fermi~Large Area
Telescope (LAT) catalogs, including the contribution from the annihilation of  dark matter in the sub-halos.  
We find that the flux sensitivity threshold strongly depends on the 
particle dark matter mass,  and more mildly also on its annihilation channel and the observation latitude. 
The results differ for the 3FGL and 2FHL catalogs, given their different energy thresholds. 
We also predict that the number of dark matter sub-halos among the unassociated sources is very small. 
A null number of detectable sub-halos in the  \Fermi-LAT 3FGL catalog would imply upper limits on the dark matter 
annihilation cross section into $b\bar{b}$ of $2\cdot 10^{-26}$ ($5\cdot 10^{-25}$) cm$^3$/s 
with $M_{\rm DM}$= 50 (1000) GeV.
We find less than one extended sub-halo in the  \Fermi-LAT 3FGL catalog. As a matter of fact, 
 the differences in the spatial and mass distribution of sub-halos between hydrodynamic and dark matter-only runs 
 do not have significant impact  on the gamma-ray dark matter phenomenology.
}

\maketitle

\section{Introduction}
\label{sec:intro}
One of the most intriguing mysteries in modern Physics is that about $85\%$ 
of all matter in the Universe is of unknown origin~\cite{Ade:2015xua}. 
Despite the extraordinary achievements in measuring the gravitational effect of this 
missing component, called dark matter (DM), 
still no direct evidence of its particle nature has been verified. 
One of the most well-motivated classes of DM particle candidates is represented 
by Weakly Interacting Massive Particles (WIMPs) (see for instance 
refs.~\cite{Bertone:2010zza,Bertone:2004pz} for a review). 
WIMPs can naturally achieve the correct relic DM abundance 
through self-annihilation in the early Universe, 
and can be searched for with several techniques. Besides direct DM 
detection experiments and searches at colliders, 
indirect DM searches aim to detect the fluxes of stable particles 
produced by DM annihilation or decay processes.
Among the possible final products of DM interactions, gamma rays are one 
of the most promising channels for DM detection, 
since they preserve the spectral and spatial features of the prompt DM signal.

Recent years have witnessed a steady progress in the field of DM indirect 
detection through gamma rays. In particular, the Large Area Telescope (LAT), 
aboard the \Fermi~satellite, is currently one of the most sensitive instruments 
collecting gamma rays from the whole sky. The \Fermi-LAT Collaboration and other 
groups have already set severe constraints on the WIMP DM parameter space with 
searches towards dwarf spheroidal galaxies \cite{Ackermann:2015zua}, of 
gamma-ray lines~\cite{Weniger:2012tx,Ackermann:2013uma}, in the diffuse Galactic 
and extragalactic emission~\cite{Calore:2013yia,Ajello:2015mfa,DiMauro:2015tfa}, 
galaxy clusters~\cite{Ackermann:2015fdi} and the Galactic 
Centre~\cite{Calore:2014xka,TheFermi-LAT:2015kwa}.

\medskip

It is well known that the sensitivity to DM detection in a specific target 
depends crucially on the distribution of DM in that particular environment. In 
the context of the concordance $\Lambda$CDM cosmology 
\cite{Bennett2013,Ade:2015xua}, a firm theoretical prediction is that structures 
in the Universe form in a hierarchical way. DM, interacting through gravity, 
collapses into structures known as DM halos \cite{Navarro1996, Navarro1997}, 
which assemble in a bottom-up way from the least massive, gradually merging to 
create larger systems \cite{White:1991mr}. These theoretical predictions are 
 confirmed by numerical simulations of structure formation modeling the 
gravitational interaction of the DM component in a full cosmological set-up 
(also known as DM only or N-body simulations), which have been widely 
successful at reproducing the large-scale distribution of structures in the 
Universe \cite{Springel2005,Boylan-Kolchin2009,Angulo2012}.

On smaller scales, i.e.~within individual DM halos, the results obtained from 
numerical simulations are more uncertain. At those scales baryon physical 
processes, that give origin to the present-day galaxy population, and that might 
also have a substantial effect on the DM distribution in halos 
\cite{2016arXiv160505323D,DiCintio2014,Pontzen2014} and its detection, are at 
play. A complete understanding of galaxy formation and evolution would require 
simulating these physical processes from first principles, but this turns out to 
be an incredibly challenging task given the extreme dynamic range of scales that 
has to be resolved. Notwithstanding these difficulties -- and the inevitable 
limitations they entail to a fully predictive theory of galaxy formation -- 
remarkable progress has been accomplished in the field over the last years. 
Hydrodynamical simulations of galaxy formation are now able to produce a galaxy 
population whose properties are in broad agreement with the observational 
constraints \cite{Vogelsberger:2014kha,Vogelsberger2014,Schaye2015}. Moreover, 
the goal of forming a disc galaxy like our own Milky Way (MW), which for decades 
has been one of the most intricate problems in the context of $\Lambda$CDM 
cosmological simulations, seems now to be achieved by many groups using 
different numerical techniques \cite{Agertz2011,Aumer2013b,Grand2016,Guedes2011,
Mollitor2015,Stinson2013,Wang2015,Colin2016,Marinacci:2013mha}.

\medskip

A robust prediction of cosmological simulations (with or without the inclusion 
of baryons) is that DM halos are populated by smaller substructures, usually 
referred to as sub-halos (SHs). The largest sample of galactic SH population 
include dwarf galaxies, which typically contain a modest amount of baryonic 
matter, i.e.~gas and stars. However, dwarfs are only the small ``visible'' 
portion of a larger population of DM SHs which lack any significant baryonic 
content and are therefore not detectable in the optical wavelength.  Besides the 
objects that are too faint to be in the reach of current optical surveys such as 
the Dark Energy Survey 
(DES)~\cite{Flaugher:2004vg,Drlica-Wagner:2015xua,Bechtol:2015cbp},  there might 
exist a number of  totally dark SHs that do not contain any star or gas. At present, 
the number of dwarf galaxies discovered in the Local Group is about 
30~\cite{Abdo:2010ex,Ackermann:2015zua,Drlica-Wagner:2015xua,Ahnen:2016qkx}. As 
DM dominated structures, SHs could emit gamma rays created by WIMPs 
self-annihilation and they may be detected as individual sources in the sky, 
depending on the signal intensity and on the astrophysical background along the 
line of sight.
On top of that, SHs that are too faint to be detected as individual sources
would instead contribute to the diffuse gamma-ray emission~\cite{Ackermann:2014usa} 
and signatures for this unresolved population of SHs might be looked 
for in the gamma-ray diffuse background intensity, e.g.~\cite{2008Natur.456...73S,Ackermann:2015tah},
and/or small scale gamma-ray anisotropies, e.g.~\cite{Calore:2014hna,Fornasa:2016ohl}.

\medskip

The \Fermi-LAT Collaboration recently released the third catalog of point sources 
(3FGL) \cite{Acero:2015gva} that contains sources detected after four years of operation 
in the energy range 0.1 -- 300 GeV with Pass 7 data.
The 3FGL catalog contains about 3000 sources, where the large majority of detected objects 
at a latitude $|b|>20^{\circ}$ are extragalactic Active Galactic Nuclei (AGN).
\Fermi-LAT also recently released a new event-level analysis, Pass 8, that increases significantly 
the acceptance of the telescope and, at the same time, improves its angular resolution~\cite{Atwood:2013rka}. 
Exploiting these improvements, the \Fermi-LAT Collaboration has compiled and released the second catalog of 
hard \Fermi-LAT sources (2FHL)~\cite{Ackermann:2015uya}. 
This catalog fills the energetic data gap with current atmospheric Cherenkov telescopes and contains 
about 360 sources detected with 80 months of exposure time and between 50 -- 2000 GeV.

In both catalogs, a large fraction of sources remain unassociated: 
about 15\% in the 2FHL and 30\% in the 3FGL. 
The probabilistic association of sources made by \Fermi-LAT takes into account 
the density of sources in the region around the gamma-ray source and its distance 
 to close-by objects detected in other wavelengths\footnote{Association using only gamma rays 
is possible only for pulsars, for which the LAT could detect the gamma-ray 
 pulsation and thus classify it as a pulsar. However, this kind of association is extremely rare.}. 
 Hence, unassociated sources are point-like gamma-ray emitters detected as such by the LAT, but
lacking association with astrophysical objects known in other wavelengths.
Interestingly, the sample of unassociated sources in the \Fermi-LAT catalogs might 
already contain  gamma-ray emitting DM SHs. Their identification requires
the determination of a realistic sensitivity flux threshold to the specific detection of DM SHs, which
is lacking in the current literature and is one of the primary goals of the present work.

\medskip

We analyze the detectability of DM SHs in current \Fermi-LAT catalogs.
Previous works have already addressed this
 issue~\cite{Bertoni:2015mla,Bertoni:2016hoh,Schoonenberg:2016aml,Hooper:2016cld}, 
examining the 3FGL source catalog and modeling the DM SHs distribution in a MW like galaxy, 
based on the N-body simulation Via Lactea II~\cite{Diemand:2008in}. The authors of~\cite{Bertoni:2015mla} 
identified 24 3FGL bright sources that may be consistent with DM (with mass about $\sim 20 - 70$ GeV) annihilation in Galactic SHs,
 as well as with faint gamma-ray pulsars. 
 In~\cite{Bertoni:2016hoh} they further scrutinized the source 3FGL J2212.5+0703 from the 
 previous subset, as a possible DM SH and gave also a plausible alternative astrophysical explanation. Both 
works set constraints on the DM annihilation cross section. 
Ref.~\cite{Schoonenberg:2016aml} updates the prior studies predicting a smaller 
 number (at most $\sim$10) of SHs which could possibly 
 be detected by the \Fermi-LAT as unassociated sources.  More recently,
  ref.~\cite{Hutten:2016jko} revisited the previous analyses focusing on the prospects of 
  detecting DM SHs with the future Cherenkov Telescope Array observatory~\cite{Acharya:2013sxa}.
Using machine learning classifiers, ref.~\cite{Mirabal:2016huj} recently 
looked for novel source classes in the sample of 3FGL unassociated sources.
They  found 34 potential candidates and 
placed upper limits on the number of Galactic SHs and, correspondingly, on the DM annihilation
cross section. Finally, the authors of~\cite{Hooper:2016cld} revisited the constraints on the DM annihilation 
cross section inferred from searches for SHs candidates among the \Fermi-LAT 3FGL unassociated sources. 
They consider the cosmological N-body simulations Via Lactea II~\cite{Diemand:2008in} and 
ELVIS~\cite{Garrison-Kimmel:2013eoa} to model the local dark matter SH population. 
Their placed limits on the DM annihilation cross section are slightly weaker than those from 
dwarfs while being stronger than those found by ref.~\cite{Schoonenberg:2016aml}. 
Our work further improves the antecedent studies with an array of novelties:
\begin{itemize}
\item The prediction of the DM SHs gamma-ray signal is based on one of the 
	most recent cosmological numerical simulations that includes baryonic 
	physics~\cite{Marinacci:2013mha,Zhu:2015jwa}. For the first time, we model the signal 
	as expected in both hydrodynamic and DM only
	simulations of the MW and we compare the results, quantifying possible differences.
\item The setups of both 3FGL and 2FHL \Fermi-LAT catalogs are simulated to derive the sensitivity of the LAT
	to DM SHs detection, the advantage being 
	a wider DM mass coverage.
\item Instead of using a fixed flux detection threshold, as usually done, we provide a realistic estimation 
	for the sensitivity of the LAT to the DM flux from SHs at high-latitude
	 as a function of DM annihilation channel, DM mass and SH Galactic latitude.
	We show that the accurate determination of the sensitivity to DM spectra leads to significant 
	differences with respect to a fixed flux threshold.
\item We estimate the detectability of extended DM SHs comparing the extension of gamma-ray 
	emission from DM interaction with the minimum extension detected in the 3FGL catalog.
\end{itemize}

We focus on the detectability of SHs as individual point sources in \Fermi-LAT
catalogs (i) for improving on previous works on this topic as explained above, and
(ii) for providing robust predictions which do not require critical extrapolations beyond the 
simulation's resolution limits (in mass and space), but rely only on simulation data instead.
Indeed, as we will see in what follows, the brightest SHs are, in general, the most massive ones.
As such, our predictions of the number of detectable SHs as individual sources depend only on
the simulation data. On the other hand, determining the distribution and luminosity function of lower-mass SHs, 
the majority of which will remain undetectable as single point sources and could contribute to the diffuse
gamma-ray background~\cite{2008Natur.456...73S}, would rely on extrapolations of the 
simulation's results beyond its mass resolution limit. This extrapolation procedure is the main theoretical 
uncertainty that affects the predictions at small scales~\cite{Sefusatti:2014vha}, and we do not tackle this 
issue down to the smallest SHs masses in the present work.
We also remind that faint (i.e. unresolved by the LAT) extragalactic sources such as blazars or Radio Galaxies are predicted to give a large contribution to the diffuse gamma-ray background (see e.g. \cite{DiMauro:2013xta,DiMauro:2013zfa}).
Nevertheless, we will discuss the effect that adding lower-mass SHs has
on our predictions.

The paper is structured as follows. In section~\ref{sec:DMmodelflux}, we describe how we 
model the DM SHs distribution -- quantifying the discrepancies between the hydrodynamic 
scenario and the pure DM one --  and their annihilation flux into gamma rays. In section~\ref{sec:sens} 
we derive the \Fermi-LAT sensitivity to DM spectra in the 3FGL and 2FHL catalogs setups. 
In section~\ref{sec:results} we present implications for DM phenomenology, namely the number 
of detectable SHs, constraints on the DM annihilation cross section, and source count distribution. 
In section~\ref{sec:extension} we discuss the possibility of resolving the extension of the detectable 
SHs. Finally, in section~\ref{sec:concl} we summarize our conclusions.

\section{Dark matter in the Galaxy and gamma-ray signals}
\label{sec:DMmodelflux}

For modeling the SH population in the Galaxy, we use the results of two 
cosmological simulations of a MW-size halo~\cite{Zhu:2015jwa}. The first 
simulation is the full hydrodynamic run Aq-C-4 in~\cite{Marinacci:2013mha} 
(``Hydro" run hereafter), while the second one is a control DM-only simulation 
of the same halo (from now on, the DMO run)~\cite{Zhu:2015jwa}. Both these 
simulations use the initial conditions of the halo C of Aquarius 
Project~\cite{Springel:2008b} (hereafter AQ08) at resolution level 4 (see 
table~\ref{tab:runs_pars} for details). While the DMO simulation models only 
gravitational interactions of the DM component, the Hydro case is 
equipped with a comprehensive galaxy formation physics model largely based on 
the Illustris simulation~\cite{Marinacci:2013mha,Vogelsberger:2014kha}. This 
model includes the most important physical processes for galaxy formation and its 
main constituents are: (i) a module for radiative cooling of the gas; (ii) a 
subgrid description of the interstellar medium and star formation out of the 
dense gas ($n\sim0.1\, {\rm cm^{-3}}$) following the prescriptions of 
\cite{Springel2003}, modified for a Chabrier \cite{Chabrier2003} initial mass 
function; (iii) routines following stellar evolution and in particular tracking 
mass and metal return from type II, type Ia supernovae and AGB stars to the 
interstellar medium; (iv) stellar feedback in the form of galactic winds 
following a kinetic implementation in which the wind velocity is scaled to the 
size of the underlying DM halo; and (v) modules for supermassive black 
hole seeding, accretion, merging and the associated AGN feedback. For reason of 
space we do not enter into the detail of the galaxy formation physics 
implementation here, but refer the reader instead to refs.~\cite{Vogelsberger2013, 
Marinacci:2013mha, Grand2016} for a full description. Both runs are performed 
with the moving-mesh code Arepo \cite{Springel2010}, a highly versatile code for 
cosmological simulations that models the hydrodynamics via a finite volume 
technique on an unstructured Voronoi mesh. This mesh is allowed to move with the 
gas, thus adapting to the flow characteristic and giving rise to a manifestly 
Galilean-invariant method that combines the strengths of both Lagrangian and 
Eulerian approaches yielding superior results in terms of accuracy. The 
evolution of the two simulated halos is followed from very high redshift 
($z=127$) down to redshift zero.

We model the SH distribution in the halo in two steps: First, we analyze the 
results of the Hydro and DMO simulations and we derive analytic 
parameterizations of the SH spatial and mass distributions. Secondly, using the 
analytic prescriptions for the statistical distribution of SH position and mass, 
we generate a mock population of Galactic SHs in multiple Monte Carlo 
realizations. In this section we describe these two steps in more detail.

\subsection{Modeling the dark matter distribution in the Galaxy}
\label{sec:model}
We consider the distribution of SHs as predicted by simulations of galaxy 
formation that include the effect of baryons in the galaxy evolution process. 
There exist three main processes driven by baryonic physics: adiabatic 
contraction, tidal disruption and reionization, which act jointly to shape the 
DM distribution in both the host halo and in its SHs. The effects of these 
processes are respectively of:  (i) increasing the density in the center of the 
Galaxy, (ii) removing both DM and luminous matter and redistribute them in the 
SHs and (iii) evaporating the gas and preventing gas accretion from the 
intergalactic medium. As a result of the baryonic actions,  usually one finds 
fewer SHs in the Hydro simulations than in the DMO ones. In particular,  fewer 
low-mass SHs are generated in the Hydro case~\cite{Zhu:2015jwa}.  Typically, 
there are also differences in the abundance and spatial distribution of the SHs, 
especially in the central region of the main halo. 
Such a depletion is caused by 
(a) gravitational shocks as SHs pass in the vicinity of the disk \cite[e.g.][]{Ostriker1972,D'Onghia2010} and (b) the 
contracted DM distribution generated by the cooling of baryons at the center of the 
halo ~\cite[e.g.][]{Blumenthal1986, Gnedin2004, Zemp2012}
As a consequence of these processes, tidal disruption is enhanced and SHs are disrupted 
more often, up to a factor of two, in the center~\cite{2009Natur.460..605D,Yurin:2014vra}.

\begin{table*}
\centering
\begin{tabular}{cccccc}
\hline
\noalign{\vskip 0.5mm}
Run & 
$R_{\rm vir}$  &
$M_{\rm tot}$  &
$m_{\rm gas}$ &
$m_{\rm DM}$  &
$\epsilon$  \\
 &$(\text{kpc})$   & 
($10^{12}$ \Msun) & 
($10^5$ \Msun) & 
($10^5$ \Msun) & 
(pc)  \\
\noalign{\vskip 0.5mm}
\hline
\noalign{\vskip 0.5mm}
DMO & 326 & 2.04  & - & 3.2 &  340 \\
\hline
\noalign{\vskip 0.5mm}
Hydro & 311 & 1.77  &  0.5 &   2.7 &  340  \\
\hline
\end{tabular}
\caption{Characteristic parameters of the two (DMO and Hydro)  simulation runs at $z = 0$. 
The virial radius $R_{\rm vir}$  is defined as a sphere enclosing an over-density of
 178 with respect to the critical density. $M_{\rm tot}$ is the
  total mass included inside $R_{\rm vir}$; $m_{\rm gas}$ and
$m_{\rm DM}$ are the mass resolution of gas and DM, respectively. Finally, $\epsilon$ is
 the gravitational softening length of the DM particles. For gas cells the softening
 length is adaptive and scaled proportionally to their sizes. Its minimum physical value is the same 
 as the one used for DM particles.}
\label{tab:runs_pars}
\end{table*}

The two simulations under study model the formation of a $2.04 \times 10^{12}$ 
\Msun~and a $1.77 \times 10^{12}$ \Msun~halo, in the DMO and Hydro case 
respectively, and of their substructures. The typical parameters of the two 
simulations are summarized in table~\ref{tab:runs_pars}. To identify the SHs we 
used the Amiga halo finder \cite{Gill2004,Knollmann2009}, a density-based algorithm which determines 
prospective SHs centers with the use of a hierarchy of adaptive grids that are 
also employed to collect the particles potentially associated to any given center. 
The final structures are then found by iteratively removing gravitationally 
unbound particles, assuming spherical symmetry, from the potential candidates 
identified in the previous step. 
We stress that this procedure is applied in the Hydro case to find all SHs of the main halo
\textit{regardless} of their stellar content. SHs identified in the Hydro simulation
can be either dark or luminous, and thus be identified as dwarf satellite galaxies. Whether or not
a SH is able to form stars depends on its mass, having that low-mass SHs are likely
to be dark, while at the high-mass end they tend to host a stellar component. The mass range for which this transition
occurs is $\sim 10^{7-8}\,{\rm M}_\odot$ \cite{Zhu:2015jwa}.

In order to avoid resolution effects, which 
may affect the properties of the SHs identified in the simulations and, 
consequently, our analysis, we apply two cuts to the sample of SHs identified by 
the halo finder. First, we consider SHs formed by at least 20 particles. Second, 
we adopt a restriction on the SHs minimum value of the maximum rotational 
velocity, $v_{\rm max}$. In both runs we require that $v_{\rm max} \gtrsim 
\sqrt{(M(< r_{\rm max}) \, G/(2.8 \, \epsilon)} \gtrsim 4$ km/s, where $G$ is 
the universal gravitational constant $G$ = 4.3 $\times 10^{-3} \rm \, pc$ 
\Msun$^{-1} \rm (km/s)^2$ and  $\epsilon$ is the gravitational softening length;
$r_{\rm max}$ is defined as the radius at which $v_{\rm max}$ is reached. 
As a result, the DMO (Hydro) run provides a reliable subsample of $\sim 1200 
~(800)$ SHs with masses $M_{\rm SH} \gtrsim  m_{\rm DM} \times 20 \sim 5.4 
\times 10^6$ \Msun. Typically, discrepancies between hydrodynamic and DMO runs 
are expected for halos with masses larger than $10^6 - 10^7$ \Msun, where stars 
can form, as  also found in ref.~\cite{Zhu:2015jwa}. However, while 
studying the impact of hydrodynamics in the mass and spatial distribution of 
Galactic SHs, we will also discuss the effect of  lower-mass SHs (see 
section~\ref{sec:results}).

\begin{figure}
	\centering
	\includegraphics[width=0.49\columnwidth]{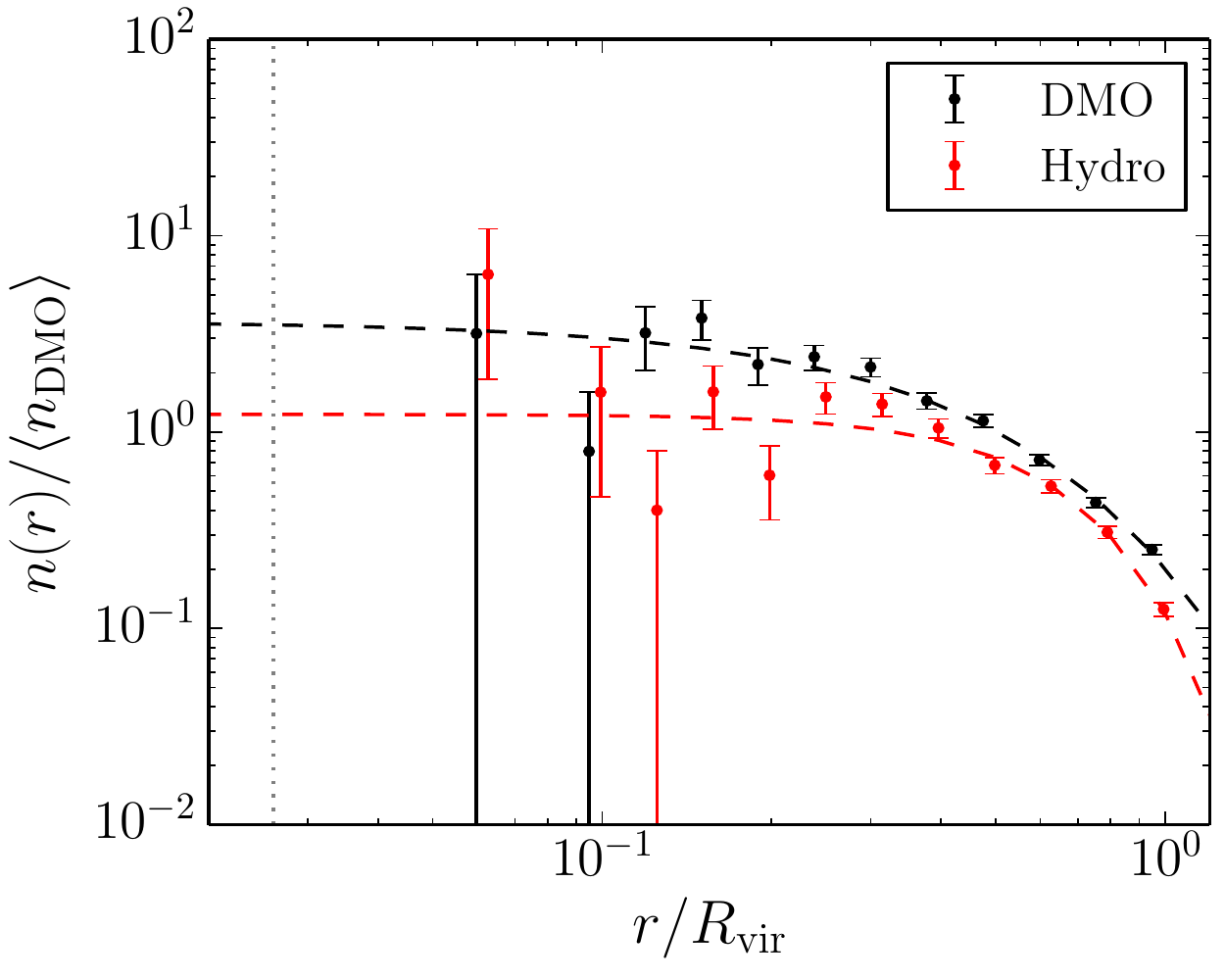}
	\includegraphics[width=0.49\columnwidth]{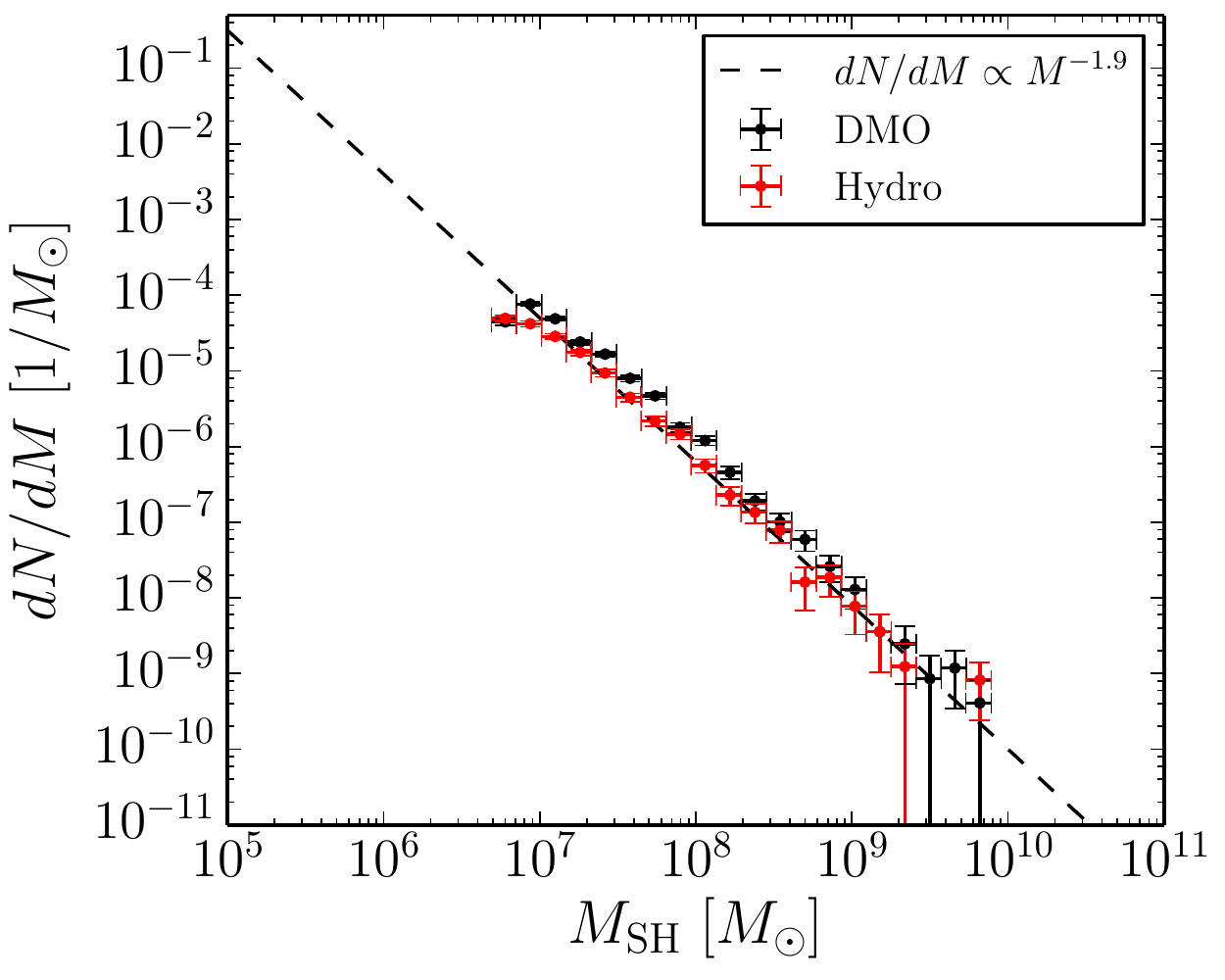}
\caption{\emph{Left panel}: Spatial distribution $n(r)$ of SHs in the Hydro (red points) and DMO
	(black points) runs~\cite{Zhu:2015jwa}, normalized to the total 
	number of SHs in the DMO run. The dashed red (black) line is the best fit for an Einasto 
	parameterization of the spatial profile, see eq.~(\ref{eq:Einastospatial}). 
	The dotted vertical line indicates the position of the Sun for the DMO run.
	\emph{Right panel}: SH differential mass abundance  $dN/dM$ in the Hydro (red points) and DMO (black points) run. 
	The lower limit of the mass axis corresponds to the smallest SH mass in AQ08 
	($M_{\rm SH}^{\rm min} = 10^5$ \Msun).
	Overlaid (dashed black curve) is the mass distribution function that best fits the AQ08 results~\cite{Springel:2008b}. }
\label{fig:SH_distributions} 
\end{figure}

\medskip 
\paragraph*{SH spatial distribution.}
From the simulations' data, we analyze the spatial distribution of SHs in the Galaxy and 
perform a fit to the radial number density of SHs $n(r)$ for both the DMO and Hydro runs 
with an Einasto function~\cite{Gao:2012tc}:
\begin{equation}
n(r)/\langle n \rangle_{\rm DMO} = n_{-2} \exp \bigg \{- \frac{2}{\alpha} \bigg[\bigg(\frac{r}{r_{-2}}\bigg)^\alpha - 1\bigg] \bigg\} \, .
\label{eq:Einastospatial}
\end{equation}
Here $r$ is the distance from the galactic center, $n(r)$ is normalized to the total number $\langle n \rangle_{\rm DMO}$ of 
SHs in the DMO run (in analogy with ref.~\cite{Zhu:2015jwa}).
The free parameters in the fit are $n_{-2}$, $\alpha$, and $r_{-2}$. 
The best-fit values that we find by minimising the $\chi^2$ are: $n_{-2} = 0.66 \pm 0.06~(0.50 \pm 0.0.03)$, 
$\alpha = 1.17 \pm 0.15~(2.20 \pm 0.29)$ and the scale radius $r_{-2}= 0.64 \pm 0.02~(0.65 \pm 0.02)$ $R_{\rm vir}$
in the DMO (Hydro) simulation, respectively. 
We show in figure~\ref{fig:SH_distributions} (left panel) the result of the fit to $n(r)$ for the DMO and Hydro runs. 
The distance $r$ is normalized to the virial radius of the main halo\footnote{$\Delta_{\rm vir}(z) = 178$ 
is the adopted virial over-density.} ($R_{\rm vir}^{\rm DMO}$ = 326 kpc and $R_{\rm vir}^{\rm Hydro}$ = 311 kpc).
As already shown in~\cite{Zhu:2015jwa}, the radial number density of SHs in the Hydro run is consistently lower than in the DMO one, 
thus meaning that the SHs are being disrupted more often in the Hydro simulation. 

\medskip
\paragraph*{SH mass distribution.}
 Most of numerical simulations in the literature (including AQ08 \cite{Springel:2008b}) have shown that 
 the SH differential mass abundance is well described by a power law $dN/dM \sim M^{-\alpha_M}$,
whose slope is slightly shallower than $-2$, over many decades in mass. 
In figure~\ref{fig:SH_distributions} (right panel) we show the number of SHs per unit mass interval, 
where the slope  of the SH mass distribution for both the DMO and Hydro runs are 
the same and consistent with AQ08 results, having $\alpha_M = 1.9$~\cite{Zhu:2015jwa}.

\medskip
\paragraph*{DM distribution and density profile of the SHs.}
The gamma-ray emissivity from DM annihilation in SHs is determined by the internal spatial profile 
of the DM SH.
Contrary to the main halo, whose DM density profile has been demonstrated to differ significantly in 
DMO and Hydro runs~\cite{Zhu:2015jwa,Schaller:2015mua,Calore:2015oya,2016arXiv160505323D}, 
in the simulations considered here the SH DM density profiles in the Hydro simulation are very similar 
to their counterparts in the DMO run~\cite{Zhu:2015jwa}.
We assume that the radial DM density profile of the SHs is described by the Einasto 
parametrization~\cite{1989A&A...223...89E}: 

\begin{equation}
\rho(r) = \rho_{s} \exp \bigg \{- \frac{2}{\alpha_\rho} \bigg[\bigg(\frac{r}{r_{s}}\bigg)^{\alpha_{\rho}} - 1\bigg] \bigg\} \, . 
\label{eq:Einasto}
\end{equation}
where $r$ is the distance from the center of the SH. 
We fix  $\alpha_\rho=0.16$, in agreement with what was found in AQ08. Therefore, the SHs density profile is 
described by a function with two free parameters: the specific density $\rho_s$ and the scale radius $r_s$, 
defined at the point where $\rho(r)$ has a slope close to a power law with index $-2$.
Given the mass of the SHs and $r_s$, $\rho_s$ is fully determined. On the other hand, $r_s$ 
has to be determined from the simulation results. 
Assuming that the density distribution of DM within each SH follows an Einasto profile, there are two quantities which 
are sufficient to determine the density profile uniquely:
the mass of the SH (or equivalently $v_{\rm max}$) and $r_{\rm max} = 2.189~r_s$~\cite{Springel:2008b}.

\begin{figure}
	\includegraphics[width=0.5\columnwidth]{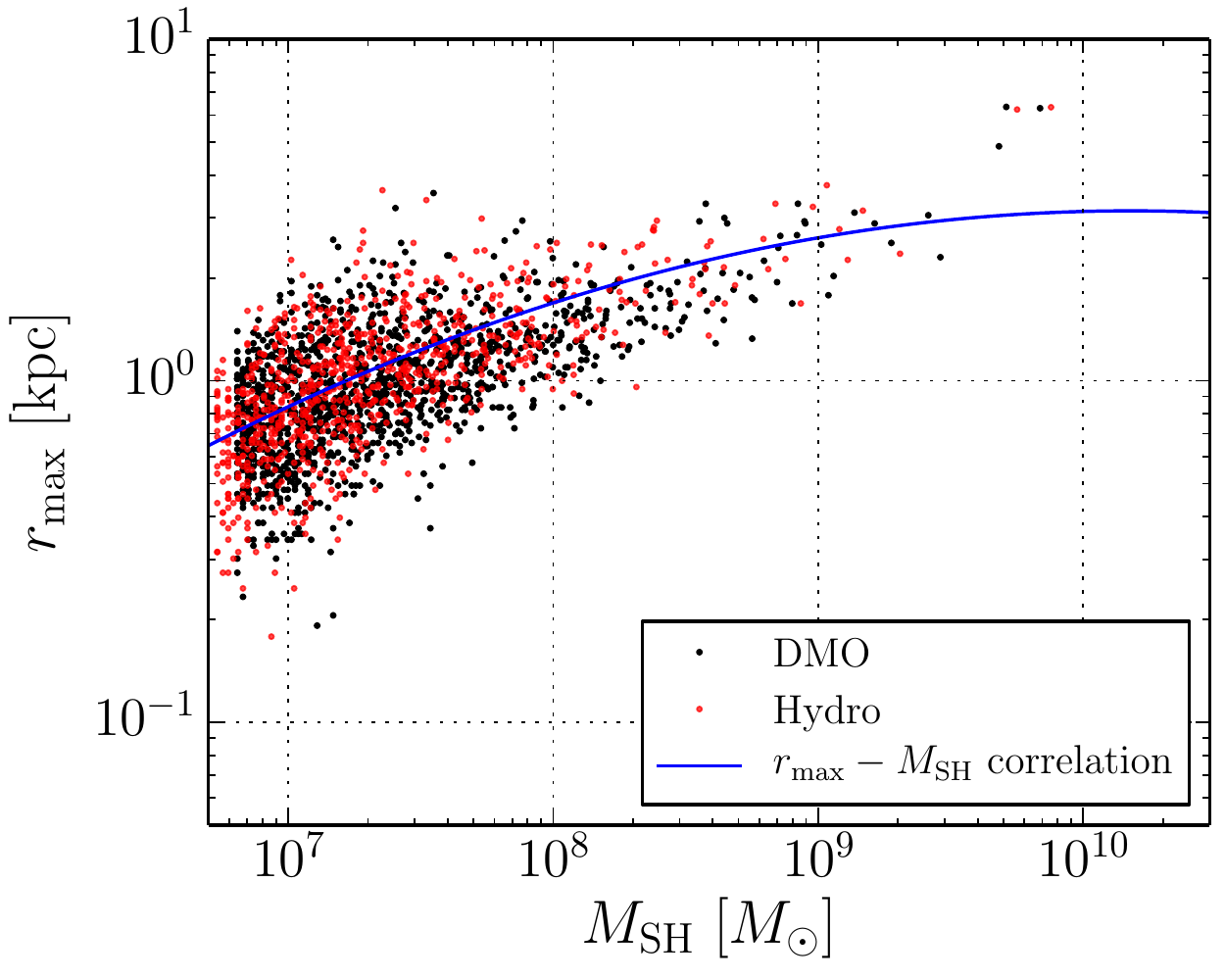}
\caption{
	$r_{\rm max}$ as a function of the SH mass, $M_{\rm SH}$  
	for the Hydro (red) and DMO (black) simulation runs. 
	Overlaid, in blue, the best-fit relation for the Hydro run as in eq. (\ref{eq:rmax}).}
\label{fig:rmaxMsh}
\end{figure}

From the simulation data, we find that the values of $r_{\rm max}$  are correlated with the SH mass.
We perform a fit to the $r_{\rm max}$ data as a function of $M_{\rm SH}$ with a polynomial function.
We derive the best-fit parameterization to be in the form: 
\begin{equation}
\log_{10}(r_{\rm max}/{\rm kpc})= a + b \,  \log_{10}(M_{\rm SH}/{\rm M}_\odot) + c\;(\log_{10}(M_{\rm SH}/{\rm M}_\odot))^2
\label{eq:rmax}
\end{equation}
with best-fit parameters\footnote{To optimize the fit, we have removed the few isolated points 
with masses $M_{\rm SH} \gtrsim 5 \times 10^9$ \Msun.}: $a = -5.384$, $b = 1.156$, 
$c = -0.056$ for the Hydro run.
The standard deviation of the data around the best-fit value is $\sigma = 0.145$. 
In figure~\ref{fig:rmaxMsh} we show $r_{\rm max}$ as a function of the 
SH mass from the DMO and Hydro runs. We overlay the corresponding
best-fit relation $r_{\rm max}$ -- $M_{\rm SH}$ for the Hydro case.
By comparing the values of $r_{\rm max}$ for the Hydro and DMO case, we can see that 
the impact of baryonic physics on the scale radius of the SHs is actually mild: the values of $r_{\rm max}$ 
in the Hydro case are quite similar to their DMO counterparts. In general, given a $M_{\rm SH}$, $r_s$ tends to be 
only slightly smaller in the DMO case.

We emphasize that the polynomial fit can be considered reliable in the parameter range
tested by the simulation, that is $5 \times 10^6 \lesssim M_{\rm SH} \lesssim 10^{10}$ \Msun. 
Its extrapolation, especially at larger masses, may be affected by sizeable statistical uncertainties.  
We note that assuming, for example, a linear distribution of $r_{\rm max}$, implies that at a given 
$M_{\rm SH}$ the $r_{\rm max}$ is larger, and
hence the scale radius $r_s$ is also larger, thus leading to different results for the DM annihilation signal.  

Indeed the scale radius, $r_s = r_{\rm max}/2.189$, which is closely 
correlated to the SH mass accretion history, affects the 
computation of the astrophysical factor (see eq.~(\ref{eq:geofactor})) appearing in the DM 
gamma-ray flux: smaller $r_s$ correspond to denser halos (see eq.~(\ref{eq:Einasto})). 

\medskip
\paragraph*{Concentration.}
A very useful parameter that can be introduced to describe the internal 
DM halos structure is the \textit{concentration}.
This quantity and its different parameterizations (in terms of SH mass, circular 
velocity and radial distance) have been widely analyzed in the literature, 
e.g.~\cite{Diemand:2008in,Pieri:2009je,Sanchez-Conde:2013yxa, Bartels:2015uba, Moline:2016pbm}.
In full generality, the concentration parameter $c$ is defined as the mean over-density 
within the radius of the peak circular velocity $r_{\rm max}$ in units of the critical density of the 
Universe at present ($\rho_{cr} = 147.897$ \Msun $/ \rm kpc^3$):

\begin{equation}
c = \frac{\bar{\rho}(r_{\rm{max}})}{\rho_{cr}} = 2\left( \frac{v_{\rm{max}}}{H_{0} \, r_{\rm{max}}}\right)^{2}\,,
\label{eq:concgen}
\end{equation}
where $H_0 = 73~ \rm km~ \rm s^{-1} \rm Mpc^{-1}$ is the Hubble constant. Equivalently, 
the concentration parameter can be cast as the ratio between the virial radius (the radius which 
encloses an average DM density $\sim 200 \times \rho_{cr} $) and the scale radius:

\begin{equation}
c = \frac{r_{\rm vir}}{r_{s}}.
\label{eq:concgen2}
\end{equation}
SHs are in general more concentrated than field 
halos of the same mass, due to the tidal force that removes material from 
their outer regions, see e.g.~\cite{2000ApJ...544..616G}.
It has also been shown that the SH concentration depends on the mass of the 
SH and on its distance from the center of the main halo~\cite{Pieri:2009je,Diemand:2008in,Sanchez-Conde:2013yxa}. 
Different concentration parameterizations depending on the SH mass and the distance have been
proposed \cite{Pieri:2009je,Diemand:2008in,Sanchez-Conde:2013yxa}, 
for both main halos and SHs. 
Nevertheless, in the present work we will not use any analytical parameterization of the concentration 
which, having been derived for other simulation results, might bias our results. Instead, we directly use the
output data of the simulation -- namely the distribution of $r_{\rm{max}}$ --  to model the scale radius. 

\medskip
\paragraph*{Monte Carlo simulation.}
Based on the modeling outlined above and derived by analyzing the simulations' data, we generate
100 Monte Carlo realizations of the SHs population in a MW-like Galaxy, for both the DMO and 
Hydro cases.
The number of simulated SHs in each realization is consistent with the total DM mass in 
the original numerical simulation~\cite{Zhu:2015jwa}. In total, we generate about 800 (1200) SHs 
in each Hydro (DMO) Monte Carlo realization. 

For each SH, we randomly extract its position in the Galaxy and its mass from the spatial and mass
distributions outlined above. We also include the uncertainty on the best-fit parameters of the distributions,
in order to  account for the the halo-to-halo variation  more realistically, i.e.~the variation
that would be present if we had disposed of more than one main host halo. 
The viral radius of each SH is defined as the tidal radius of the SH, modeled 
according to eq.~(12) of AQ08, and dependent on the SH position and mass.

In the Monte Carlo simulation, given the SH mass, we compute the value of $r_{\rm max}$ from the polynomial 
best-fit and we add a 3$\sigma$ log-normal dispersion about the best-fit relation in figure~\ref{fig:rmaxMsh}. 
We then get the value of $r_s = r_{\rm max}/2.189$~\cite{Springel:2008b}.

\begin{figure}
	\centering
	\includegraphics[width=0.5\columnwidth]{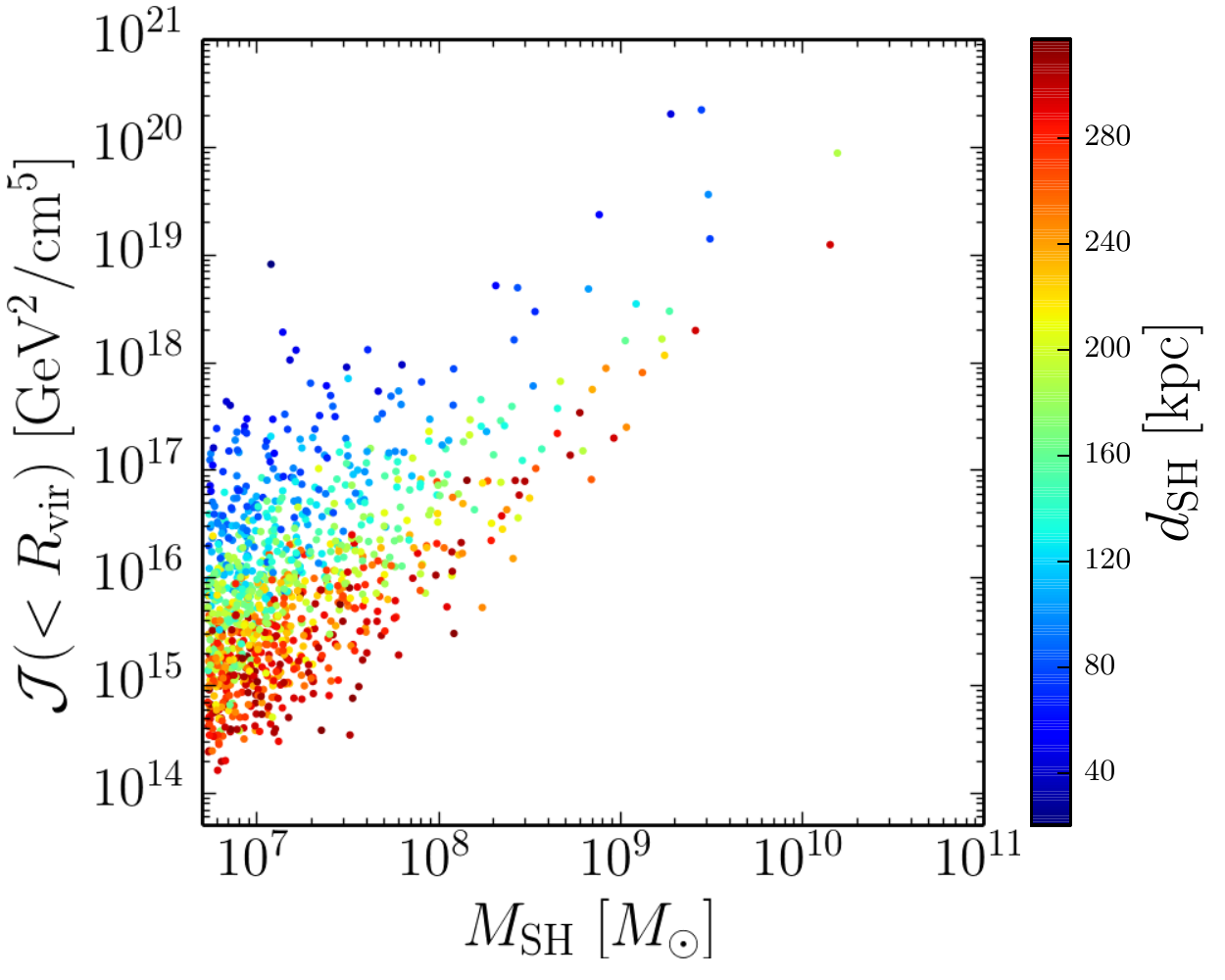}
\caption{Scatter plot of \Jf~values, $\mathcal{J}$, 
	as a function of the SH mass, $M_{\rm SH}$  in one Hydro realization of our Monte Carlo simulation. 
	The color-bar represents the distance  of the SH from Earth, hereafter $d_{\rm SH}$. }
\label{fig:Rs_Jf_compare}
\end{figure}

\subsection{Dark matter annihilation gamma-ray signatures}
\label{sec:prediction}

DM particle annihilation produces gamma rays through direct emission, 
the so-called prompt mechanism, and through indirect processes, such as the Inverse 
Compton scattering of final electrons and positrons with low-energy ambient photons,  
or bremsstrahlung of the same population of electrons and positrons with the interstellar gas.
Usually different primary annihilation channels are studied assuming 
a branching ratio of 100\% in each channel separately. 
Here, we take into account one typical hadronic annihilation channel, $b\bar{b}$, 
and the leptonic channel that gives the largest DM gamma-ray flux, i.e.~$\tau^+\tau^-$. 
For  both pairs, the most important gamma-ray emission mechanism is the 
prompt one \cite{Calore:2013yia,Cirelli:2010xx}. We therefore do not include any secondary emission in this analysis.

The flux of photons, $F$, integrated over the energy range $\Delta E = E_1 - E_0$ from a given region of the sky
and produced by the annihilation of self-conjugated DM particles is calculated as:
\begin{equation}
F_{[E_1,E_0]} = \frac{ \sv }{8\pi M^2_{\rm{DM}}} \; \mathcal{I}_{[E_1,E_0]} \; \mathcal{J},
\label{eq:phflux}
\end{equation}
where $M_{\rm{DM}}$ is the DM particle mass, $\sv$ is the thermal averaged annihilation cross section, 
$\mathcal{I}_{[E_1,E_0]}$ is the integrated energy spectrum
$\mathcal{I}_{[E_1,E_0]} = \int_{E_0}^{E_1} dN_{\rm{DM}}/dE \;dE$  
in the energy range $[E_0,E_1]$. The energy spectra of gamma rays produced 
from DM annihilation in $b\bar{b}$ and $\tau^+\tau^-$ 
channels are taken from~\cite{Cirelli:2010xx}, where they are calculated using 
PYTHIA 8~\cite{Sjostrand:2014zea} event generator.
Finally, $\mathcal{J}$ is the geometrical factor defined as:
\begin{equation}
\label{eq:geofactor}
\mathcal{J} = 2 \pi \int_{\theta_{\rm{min}}}^{\theta_{\rm{max}}} d\theta \sin{(\theta)} \int_{\rm{l.o.s}} \rho^2(r(l,\theta)) dl ,
\end{equation}
where $\theta$ is the opening angle with respect to the line of sight $l$ that points to 
the center of the SH; $\theta_{\rm{min}}$ is set to 0 and thus corresponds to the direction 
of the SH center, while $\theta_{\rm{max}}$ is $\pi$. 
The radial distance $r$ from the center of the SH  is defined as
$r^2 = d^2 + l^2 -2\,l\,d\,\cos{(\theta)}$. 

The $\mathcal{J}$-factor encodes the information about the geometry of the emission and it is a direct measure of 
the intensity of the signal, being $F_{[E_1,E_0]} \propto \mathcal{J}$.
We compute the \Jf~for the two sets of SHs in our Monte Carlo simulations. In figure~\ref{fig:Rs_Jf_compare} we show  
the values of the \Jf~versus the SH mass for the Hydro case. The color code indicates the distance of the SH from Earth (in kpc),
assuming the Sun distance from the Galactic center to be 8.5 kpc --
 blue being the solutions for the closest SHs and red those for the farthest ones. We have proven that 
 the results for the Hydro and DMO cases are fully comparable, as it can already been deduced from figure~\ref{fig:rmaxMsh}.
Given the mild difference between the Hydro and the DMO cases, in the following we will show results only for the Hydro case.

\bigskip

Another important ingredient for the determination of the DM annihilation gamma-ray
signal is the spectral energy distribution of the signal, the $dN_{\rm DM}/dE$.
We will provide in the next section the flux sensitivity to detect a DM SH as a 
function of the DM channel, mass and Galactic latitude. This result is derived simulating 
the gamma-ray flux from DM SHs and analyzing the simulations with {\it Fermi}-LAT 
Science Tools in order to find significance of their emission. It is thus useful to model 
the emission from DM annihilation with a spectral shape already included in the Science 
Tools. Among all the possible functions (see Science 
Tools\footnote{\url{http://fermi.gsfc.nasa.gov/ssc/data/analysis/scitools/source_models.html}}) 
the more flexible is the so called 
super-exponential cutoff parameterization, given by the following equation \cite{Acero:2015gva}:
\begin{equation}
\frac{dN_{\rm{DM}}}{dE} (E[\rm{MeV}])= K \left(\frac{E}{E_0}\right)^{-\Gamma} \exp{\left( - \left(\frac{E}{E_{\rm{cut}}}\right)^{\beta} \right)},
\label{eq:supexp}
\end{equation}
where $E_0=10^{3}$ MeV is the pivot energy, $\Gamma$ is the spectral 
index, $E_{\rm{cut}}$ is the energy cutoff and $\beta$ is the curvature index.
Depending on the DM mass, $\Gamma=[0.90,0.10]$ and the spectrum has an exponential 
cutoff after the peak, which is located at an energy of about $E_{\rm{peak}}=M_{\rm{DM}}/20$ 
for $b\bar{b}$ channel, and  $E_{\rm{peak}}=M_{\rm{DM}}/3$ for $\tau^+\tau^-$ channel. 
We perform a fit to the DM annihilation gamma-ray spectra taken from~\cite{Cirelli:2010xx} 
using eq.~(\ref{eq:supexp}).
This functional form provides a very good fit to DM spectra for all DM masses between $8$ 
and $10^5$ GeV. The values of the best-fit parameters are reported in table~\ref{tab:bbspectra} 
for both $b\bar{b}$ and $\tau^+\tau^-$ DM annihilation channels.

\begin{figure}
	\centering
	\includegraphics[width=0.49\columnwidth]{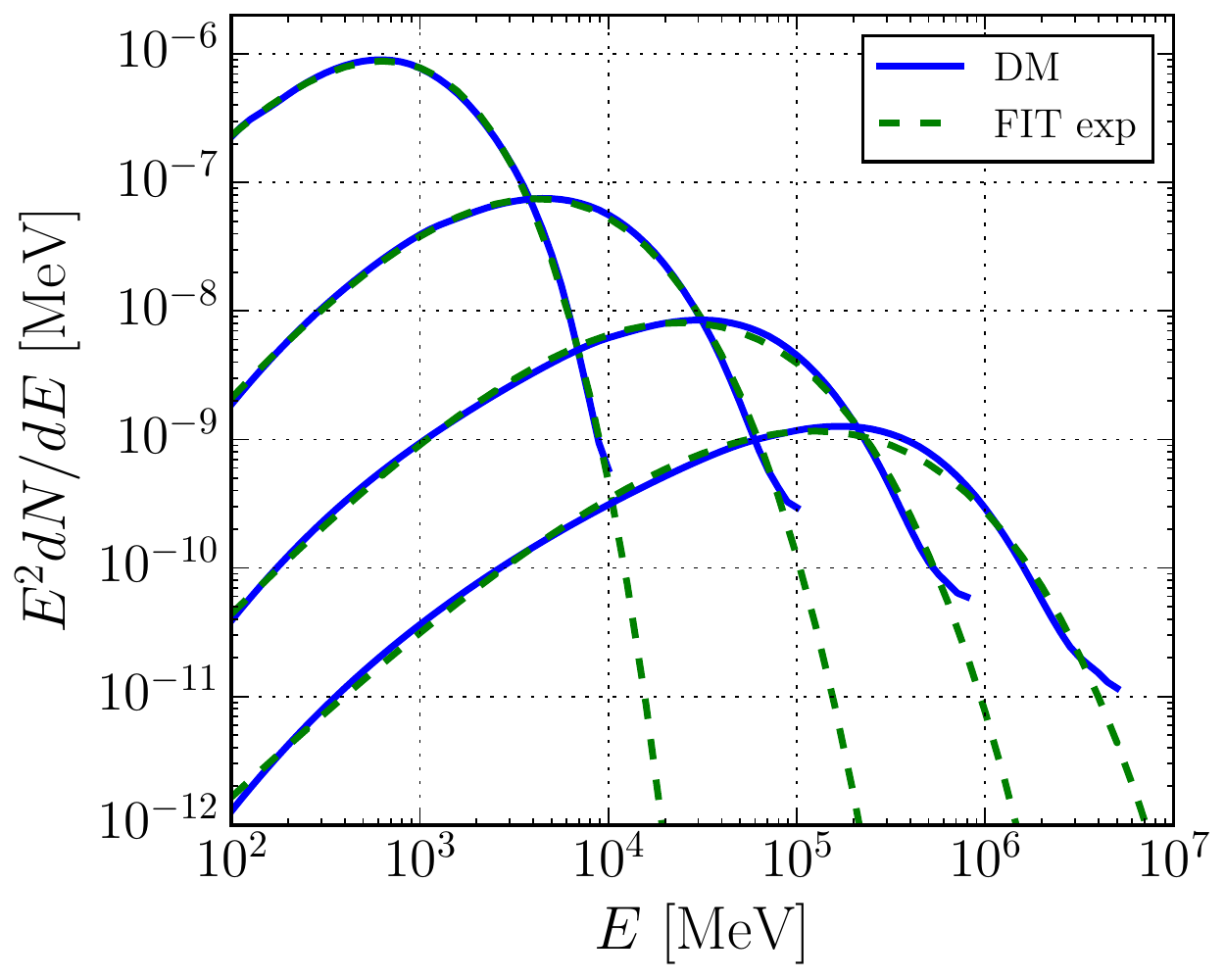}
	\includegraphics[width=0.49\columnwidth]{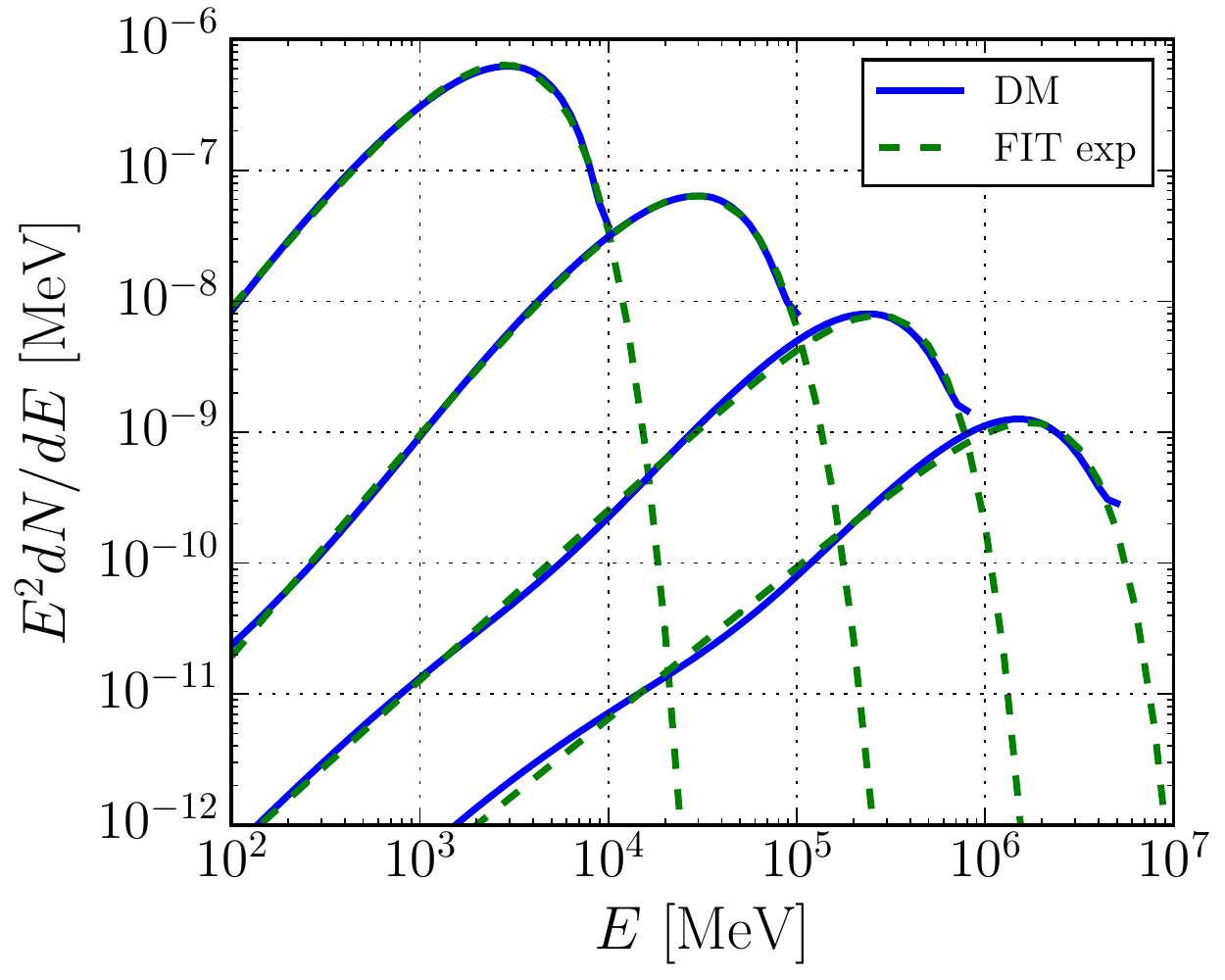}
\caption{Fit to DM annihilation gamma-ray spectra from ref.~\cite{Cirelli:2010xx} 
	with the super-exponential cutoff function (eq.~(\ref{eq:supexp})), for 
	$M_{\rm DM}$ = 10, 100, 800, 5000 GeV (curves from left to right).
	In the left (right) panel a $b\bar{b}$ ($\tau^+$$\tau^-$) annihilation channel is assumed.
	The spectra from ref.~\cite{Cirelli:2010xx} are normalized, dividing by the factor $(8 \times \pi \times M_{\rm DM}^2)$.
	}
\label{fig:dmspectrafit} 
\end{figure}

\begin{table}[t]
\center
\begin{tabular}{|c|c|c|c|c||c|c|c|c|}
\hline
$M_{\rm{DM}}$	&  $K$  &  $\Gamma$  &      $E_{\rm{cut}}$   &  $\beta$ &  $K$  &  $\Gamma$  &      $E_{\rm{cut}}$   &  $\beta$ \\
\hline
\hline
8   &   9.735$\cdot10^{-11}$  &  0.096  &  7.294$\cdot10^{1}$  &  0.594  &   8.491$\cdot10^{-13}$  &  0.303  &  1.676$\cdot10^{3}$  &  1.210  \\  
\hline
10   &   4.989$\cdot10^{-11}$  &  0.143  &  8.624$\cdot10^{1}$  &  0.581  &   4.833$\cdot10^{-13}$  &  0.280  &  1.996$\cdot10^{3}$  &  1.170  \\  
\hline
15   &   3.435$\cdot10^{-11}$  &  0.000  &  6.617$\cdot10^{1}$  &  0.520  &   1.738$\cdot10^{-13}$  &  0.223  &  2.653$\cdot10^{3}$  &  1.093  \\  
\hline
20   &   1.817$\cdot10^{-11}$  &  0.000  &  6.908$\cdot10^{1}$  &  0.498  &   8.150$\cdot10^{-14}$  &   0.200  &  3.313$\cdot10^{3}$  &  1.054  \\  
\hline
30   &   7.806$\cdot10^{-12}$  &  0.000  &  7.021$\cdot10^{1}$  &  0.468  &   2.691$\cdot10^{-14}$  &   0.197  &  4.806$\cdot10^{3}$  &  1.028  \\  
\hline
40   &   4.109$\cdot10^{-12}$  &  0.000  &  7.455$\cdot10^{1}$  &  0.452  &   1.211$\cdot10^{-14}$  &   0.210  &  6.484$\cdot10^{3}$  &  1.026  \\  
\hline
50   &   2.674$\cdot10^{-12}$  &  0.000  &  7.295$\cdot10^{1}$  &  0.437  &   6.561$\cdot10^{-15}$  &   0.221  &  8.187$\cdot10^{3}$  &  1.025  \\  
\hline
60   &   1.802$\cdot10^{-12}$  &  0.000  &  7.487$\cdot10^{1}$  &  0.427  &   3.960$\cdot10^{-15}$  &   0.239  &  1.018$\cdot10^{4}$  &  1.036  \\  
\hline
80   &   9.890$\cdot10^{-13}$  &  0.000  &  7.605$\cdot10^{1}$  &  0.412  &   1.808$\cdot10^{-15}$  &   0.273  &  1.456$\cdot10^{4}$  &  1.063  \\  
\hline
100   &   6.552$\cdot10^{-13}$  & 0.000  &  7.196$\cdot10^{1}$  &  0.398  &   9.964$\cdot10^{-16}$  &   0.290  &  1.880$\cdot10^{4}$  &  1.074  \\  
\hline
150   &   2.960$\cdot10^{-13}$  & 0.000  &  6.834$\cdot10^{1}$  &  0.376  &   3.557$\cdot10^{-16}$  &   0.363  &  3.281$\cdot10^{4}$  &  1.146  \\  
\hline
200   &   1.669$\cdot10^{-13}$  & 0.009  &  6.613$\cdot10^{1}$  &  0.362  &   1.861$\cdot10^{-16}$  &   0.437  &  5.035$\cdot10^{4}$  &  1.226  \\  
\hline
300   &   4.641$\cdot10^{-14}$  & 0.142  &  1.279$\cdot10^{2}$  &  0.368  &   8.019$\cdot10^{-17}$  &   0.528  &  8.753$\cdot10^{5}$  &  1.321  \\  
\hline
400   &   2.098$\cdot10^{-14}$  & 0.209  &  1.842$\cdot10^{2}$  &  0.369  &   4.603$\cdot10^{-17}$  &   0.589  &  1.284$\cdot10^{5}$  &  1.393  \\  
\hline
500   &   1.134$\cdot10^{-14}$  & 0.269  &  2.560$\cdot10^{2}$  &  0.371  &   3.013$\cdot10^{-17}$  &   0.627  &  1.688$\cdot10^{5}$  &  1.431  \\  
\hline
600   &   7.073$\cdot10^{-15}$  &  0.305  &  3.222$\cdot10^{2}$  &  0.372  &   2.154$\cdot10^{-17}$  &   0.658  &  2.113$\cdot10^{5}$  &  1.468  \\  
\hline
800   &   3.685$\cdot10^{-15}$  &  0.337  &  4.000$\cdot10^{2}$  &  0.370  &   1.273$\cdot10^{-17}$  &   0.698  &  2.965$\cdot10^{5}$  &  1.509  \\  
\hline
1000   &   2.034$\cdot10^{-15}$  & 0.397  &  5.907$\cdot10^{2}$  &  0.372  &   8.507$\cdot10^{-18}$  &   0.726  &  3.823$\cdot10^{5}$  &  1.533  \\  
\hline
1500   &   8.566$\cdot10^{-16}$  &  0.431  &  7.715$\cdot10^{2}$  &  0.364  &   4.081$\cdot10^{-18}$  &   0.766  &  5.952$\cdot10^{5}$  &  1.555  \\  
\hline
2000   &   4.796$\cdot10^{-16}$  &  0.444  &  8.698$\cdot10^{2}$  &  0.356  &   2.407$\cdot10^{-18}$  &   0.787  &  8.052$\cdot10^{5}$  &  1.558  \\  
\hline
3000   &   1.995$\cdot10^{-16}$  &  0.491  &  1.273$\cdot10^{3}$  &  0.353  &   1.149$\cdot10^{-18}$  &   0.814  &  1.228$\cdot10^{6}$  &  1.559  \\  
\hline
4000   &   1.155$\cdot10^{-16}$  &  0.494  &  1.336$\cdot10^{3}$  &  0.343  &   6.703$\cdot10^{-19}$  &   0.827  &  1.632$\cdot10^{6}$  &  1.535  \\  
\hline
5000   &   7.032$\cdot10^{-17}$  &  0.530  &  1.822$\cdot10^{3}$  &  0.345  &   4.473$\cdot10^{-19}$  &   0.839  &  2.053$\cdot10^{6}$  &  1.531  \\  
\hline
6000   &   5.035$\cdot10^{-17}$  &  0.527  &  1.811$\cdot10^{3}$  &  0.337  &   3.160$\cdot10^{-19}$  &   0.844  &  2.444$\cdot10^{6}$  &  1.508  \\  
\hline
8000   &   2.944$\cdot10^{-17}$  &  0.526  &  1.845$\cdot10^{3}$  &  0.327  &   1.801$\cdot10^{-19}$  &   0.849  &  3.184$\cdot10^{6}$  &  1.461  \\  
\hline
10000   &   1.826$\cdot10^{-17}$  &  0.557  &  2.488$\cdot10^{3}$  &  0.329  &   1.185$\cdot10^{-19}$  &   0.856  &  3.966$\cdot10^{6}$  &  1.445  \\  
\hline
15000   &   8.692$\cdot10^{-18}$  &  0.554  &  2.477$\cdot10^{3}$  &  0.314  &   5.408$\cdot10^{-20}$  &   0.863  &  5.759$\cdot10^{6}$  &  1.380  \\  
\hline
20000   &   5.204$\cdot10^{-18}$  &  0.545  &  2.303$\cdot10^{3}$  &  0.303  &   3.033$\cdot10^{-20}$  &   0.864  &  7.370$\cdot10^{6}$  &  1.315  \\  
\hline
30000   &   2.394$\cdot10^{-18}$  &  0.562  &  2.745$\cdot10^{3}$  &  0.295  &   1.359$\cdot10^{-20}$  &   0.866  &  1.047$\cdot10^{7}$  &  1.235  \\  
\hline
50000   &   9.194$\cdot10^{-19}$  &  0.574  &  3.126$\cdot10^{3}$  &  0.284  &   5.003$\cdot10^{-21}$  &   0.870  &  1.581$\cdot10^{7}$  &  1.111  \\  
\hline
100000   &   3.213$\cdot10^{-18}$  &  0.138  &  1.230  &  0.172  &   1.910$\cdot10^{-20}$  &   0.359  &  1.000  &  0.133  \\  
\hline
\hline
\hline
\end{tabular}
\caption{Values of the parameters $K$ (in MeV$^{-1}$), $\Gamma$, $E_{\rm{cut}}$ (in MeV), 
	and $\beta$ entering the super-exponential cutoff	function eq.~(\ref{eq:supexp}), 
	from a fit to the gamma-ray spectra from DM annihilation, for
	$b\bar{b}$ (columns from 2 to 5) and $\tau^+\tau^-$ channels (columns from 6 to 9) at 
	given DM mass $M_{\rm{DM}}$ (in GeV).}
\label{tab:bbspectra}
\end{table}

\section{\Fermi-LAT sensitivity to dark matter spectra}
\label{sec:sens}
The main aim  of this paper is to predict the detectability of Galactic DM SHs,
modeled according the latest hydrodynamic simulations, by the \Fermi-LAT. 
At this scope we implement, for the first time, the characteristics of both the  
low-energy 3FGL ($E>0.1$ GeV) and the high-energy 
2FHL ($E>50$ GeV)  \Fermi-LAT catalogs. 
One of the main novelties of this paper is the realistic estimation of the flux sensitivity of \Fermi-LAT to DM SHs detection.
The flux sensitivity is defined as the flux at which the Test Statistic (TS)\footnote{The TS is 
defined as TS=$2( \log{\mathcal{L}(\mu_k)} -\log{\mathcal{L}(0)} )$ where $\mathcal{L}(\mu_k)$ 
is the likelihood for the presence of the source (the spectrum of the source depends on generic 
parameters $\mu_k$) and $\mathcal{L}(0)$ is the likelihood of the null hypothesis of background 
only emission (by the interstellar and isotropic emission).} 
for the SH detection is equal to 25. This is the typical TS value adopted in the \Fermi-LAT catalogs 
to claim the detection of sources.
Previous works have assumed a fixed threshold to determine the detection of SHs
(see e.g. \cite{Bertoni:2015mla,Schoonenberg:2016aml}). 
In this work, we show how the sensitivity flux depends on the DM annihilation channel, DM mass and 
position of the SH in the sky.
The assumption of a fixed sensitivity threshold could turn out to be
not accurate enough for the following reasons:
\begin{itemize}
\item 
	The spectral representations of sources in \Fermi-LAT catalog are 
	energy power laws with spectral index $\Gamma$ ($dN/dE \propto E^{-\Gamma}$), 
	or suitable modifications for correcting  curved or  exponentially cut-off spectra.
	The LAT, as shown in \cite{Acero:2015gva}, has a strong bias for the
	detection of sources 
	with a given flux  as a function of the spectral index. Indeed, the telescope detects more easily
	lower photon fluxes for sources with harder spectra. This bias could be alleviated considering 
	energy fluxes ($S=\int_{0.1\rm{GeV}} dN/dE \, E \, dE$) instead of photon fluxes above 
	100 MeV ($F=\int_{0.1\rm{GeV}} dN/dE \, dE$),
	as done in \cite{Schoonenberg:2016aml}, or considering 
	photon fluxes integrated above 1 GeV, as in \cite{Bertoni:2015mla}.
	However, even when considering photon fluxes for $E>1$ GeV or energy fluxes, a
	 dependence on the spectrum assumed for the source
	still remains and, as a consequence, the sensitivity threshold might vary up to a factor 
	of 2 \cite{Acero:2015gva}.
	By assuming a fixed sensitivity threshold, the dependence of the sensitivity itself
	 on the specific source spectrum is ignored, hence leading to possible biases. 
	 \item Both the angular resolution and the acceptance of the LAT strongly depend on energy. 
	The angular resolution, for example, is a factor 5 better at 1 GeV than at 100 MeV. 
	This is quite relevant for the detection of DM SHs, since the shape of the DM 
	annihilation gamma-ray energy 
	spectrum changes significantly as a function of the annihilation channel and DM mass. 
	For example, the peak of the spectrum for a DM candidate annihilating into $b\bar{b}$ and 
	with a mass $M_{\rm{DM}}=10$ GeV is at a few hundreds MeV, while
	 for $M_{\rm{DM}}=100$ GeV the peak 
	 appears at few GeV. 
        Indeed, as we will show in the next sections, there is a strong dependence of the sensitivity 
        on the DM particle mass.
\item AGN are the most numerous source population detected by the LAT and the estimation of the 
	sensitivity flux from \Fermi-LAT catalogs is thus mostly related to the gamma-ray spectrum of 
	these objects. However, DM gamma-ray spectra are very different from the spectral energy distribution of AGN. 
	Most of AGN spectra are modeled in the 3FGL with a power law spectra with an average index of about $\Gamma = 2.4$ 
         while, as shown in section~\ref{sec:prediction}, the DM spectrum can be well parametrized
	 by a super-exponential cutoff.  Therefore, assuming a fixed sensitivity threshold for DM SHs 
	 detection based on the sources in the 3FGL and 2FHL catalogs further ignores the dependence on the spectral shape of the signal.  
\end{itemize}

\noindent
In this section we present the method that we have developed to estimate the 
flux sensitivity of \Fermi-LAT to DM SH gamma-ray spectra.
We start fully simulating the gamma-ray sky, including the interstellar and isotropic emissions.
Then, we simulate DM SHs with different DM masses, both for $b\bar{b}$ and $\tau^+ \tau^-$ 
annihilation channels. We also consider different positions of SHs in the sky, by positioning them 
at different Galactic latitudes, $b$. We neglect the longitude dependence of the sensitivity flux 
because at high latitudes ($|b|>20^{\circ}$) the longitudinal variations of the background emissions 
are negligible compared to the changes induced by variations of the Galactic latitude.
All-sky gamma-ray maps are created for the same exposure times, energy range and
 instrument response functions of the two adopted catalogs.
Implementing the sensitivity also for the 2FHL catalog (beside the 3FGL one) is motivated by 
the fact that this is the first \Fermi-LAT source catalog made with the new Pass 8 event selection. 
Given the significant improvement of this new dataset, we can provide precise predictions for 
the detection of DM SHs in an energy range that will be of particular interest for the future Cherenkov 
Telescope Array observatory~\cite{Acharya:2013sxa} (see also ref.~\cite{Hutten:2016jko}).

Operationally, we generate gamma-ray maps of the emission of DM SHs at different latitudes and 
for different DM channel and masses. We then run the typical detection pipeline in the Binned 
Likelihood case of the \Fermi-LAT Science 
Tools\footnote{\url{http://fermi.gsfc.nasa.gov/ssc/data/analysis/scitools/binned_likelihood_tutorial.html}}, which 
includes running the $gtselect$, $gtmktime$, $gtbin$, $gtsrcmap$ and finally $gtlike$ tools.
For each DM annihilation channel, DM mass and latitude we derive the flux for which TS=25:
this represents the sensitivity flux for that particular DM SH configuration.
We note that the uncertainty on the SH flux threshold (also for very bright SHs) depends 
on the specific run of the \Fermi-LAT Science Tools and can vary within a factor of about $20\%$.

In the next two sections we show the results for the sensitivity flux for the 3FGL and 2FHL catalog setups. 

\subsection{Sensitivity to dark matter fluxes for the 3FGL catalog setup}
In the case of the 3FGL catalog setup we consider 4 years of data (from 2008 August 4, to 2012 July 3) 
and the energy range $0.1-300$ GeV. As done in the 3FGL catalog, in order to reduce the contamination 
from the Earth limb, events with zenith angles larger than $100^{\circ}$ are excised.
We simulate the interstellar emission model (IEM) using {\tt gll\_iem\_v05\_rev1.fit}, the isotropic template 
using {\tt iso\_source\_v05.txt} and DM SHs at different longitudes and latitudes in the sky. We vary the 
DM mass between 8 GeV and 10 TeV. 
We adopt 10 logarithmic bins and a region of interest (ROI) 
with a radius of 15$^{\circ}$ around each DM SH, dividing it in spatial bin with size of $0.2 \times 0.2$ deg$^2$.
 We aim at  finding a DM SH flux for different position in the Galaxy, DM mass and annihilation channel, 
 and  derive the relation between this flux and the SH TS.

\begin{figure}
	\centering
	\includegraphics[width=0.49\columnwidth]{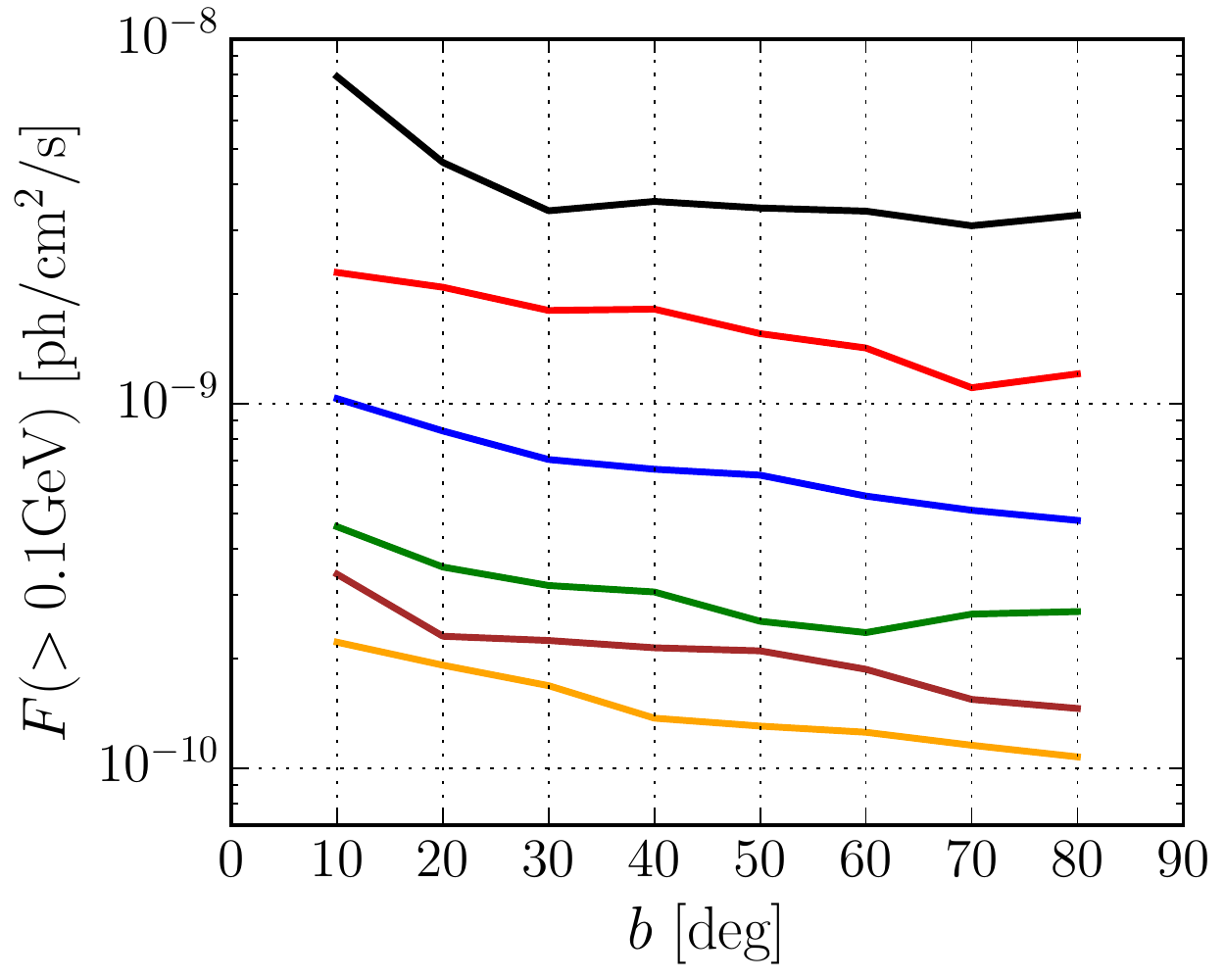}
	\includegraphics[width=0.49\columnwidth]{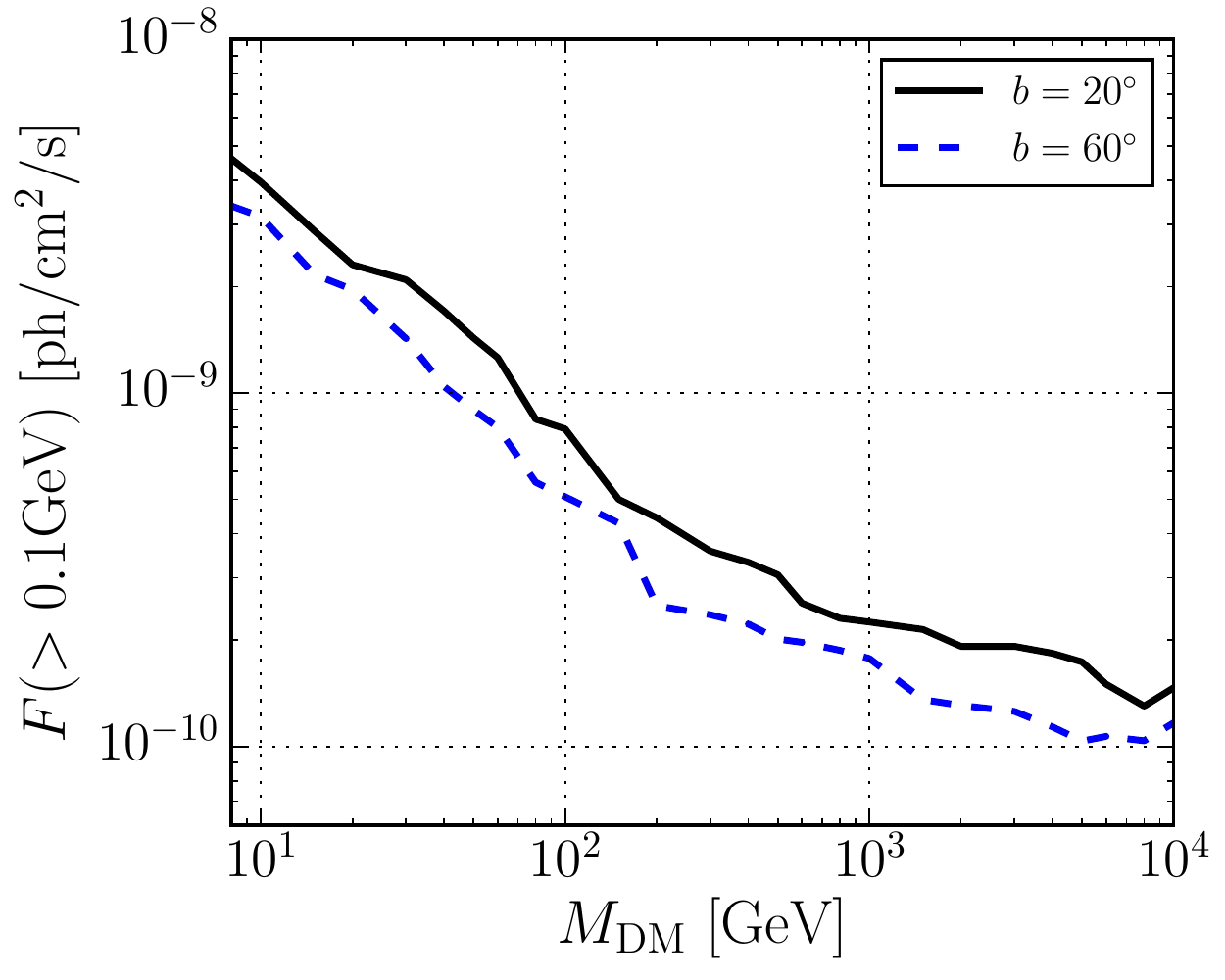}
\caption{Flux sensitivity threshold of \Fermi-LAT 3FGL to DM annihilation spectra for $b \bar{b}$ 
	annihilation channel.
	\emph{Left panel}: Flux sensitivity threshold as a function of position (latitude) of the SH for, from top to bottom,
	$M_{\rm{DM}}=$ 8 (black), 30 (red), 80 (blue), 300 (green), 600 (brown), 1200 (orange) GeV.
	\emph{Right panel}: Flux sensitivity threshold as a function of DM mass for $b=20^{\circ}$ and $60^{\circ}$ of the SH.
	} 
\label{fig:sens3FGLbb} 
\end{figure}

\begin{figure}
	\centering
	\includegraphics[width=0.49\columnwidth]{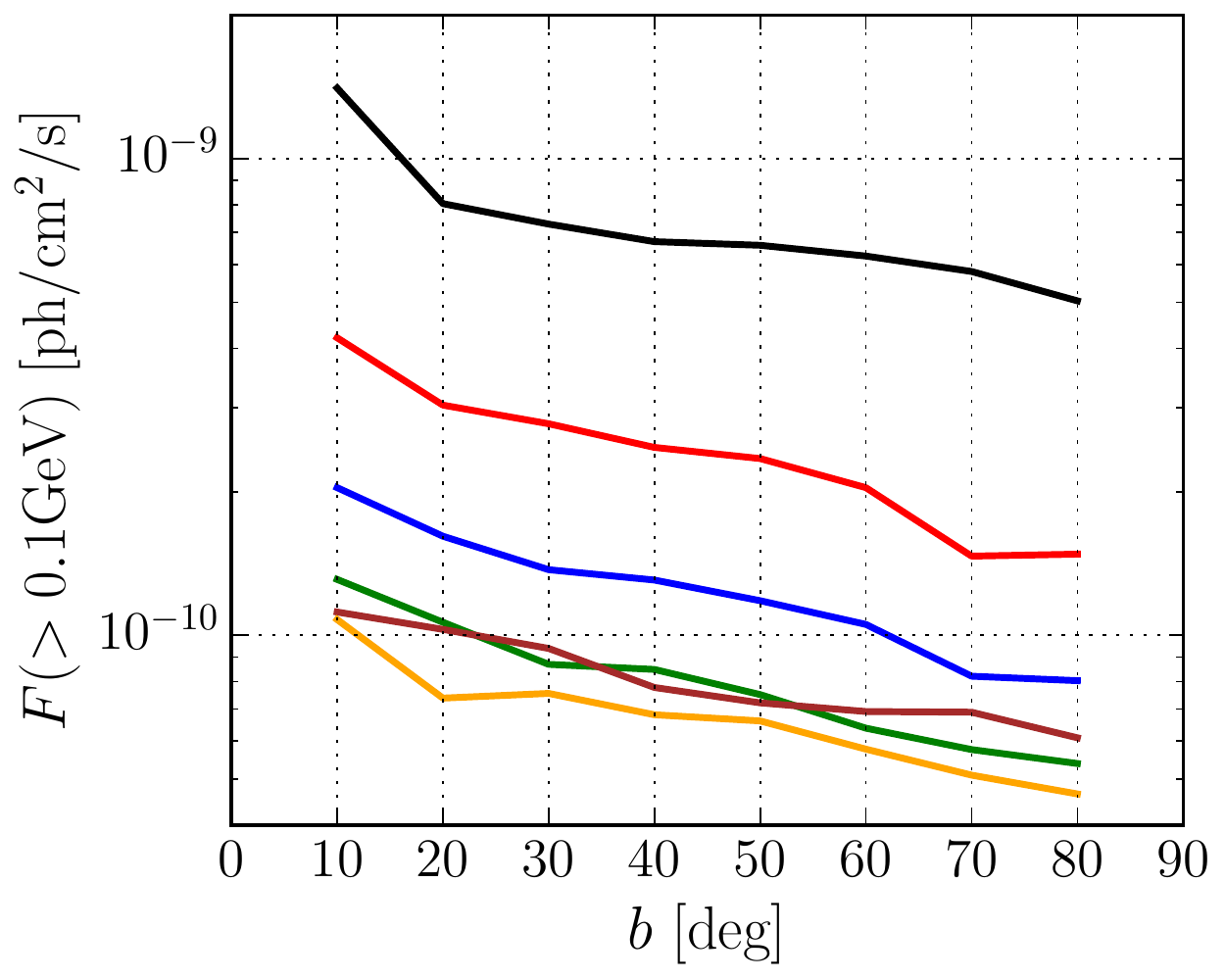}
	\includegraphics[width=0.49\columnwidth]{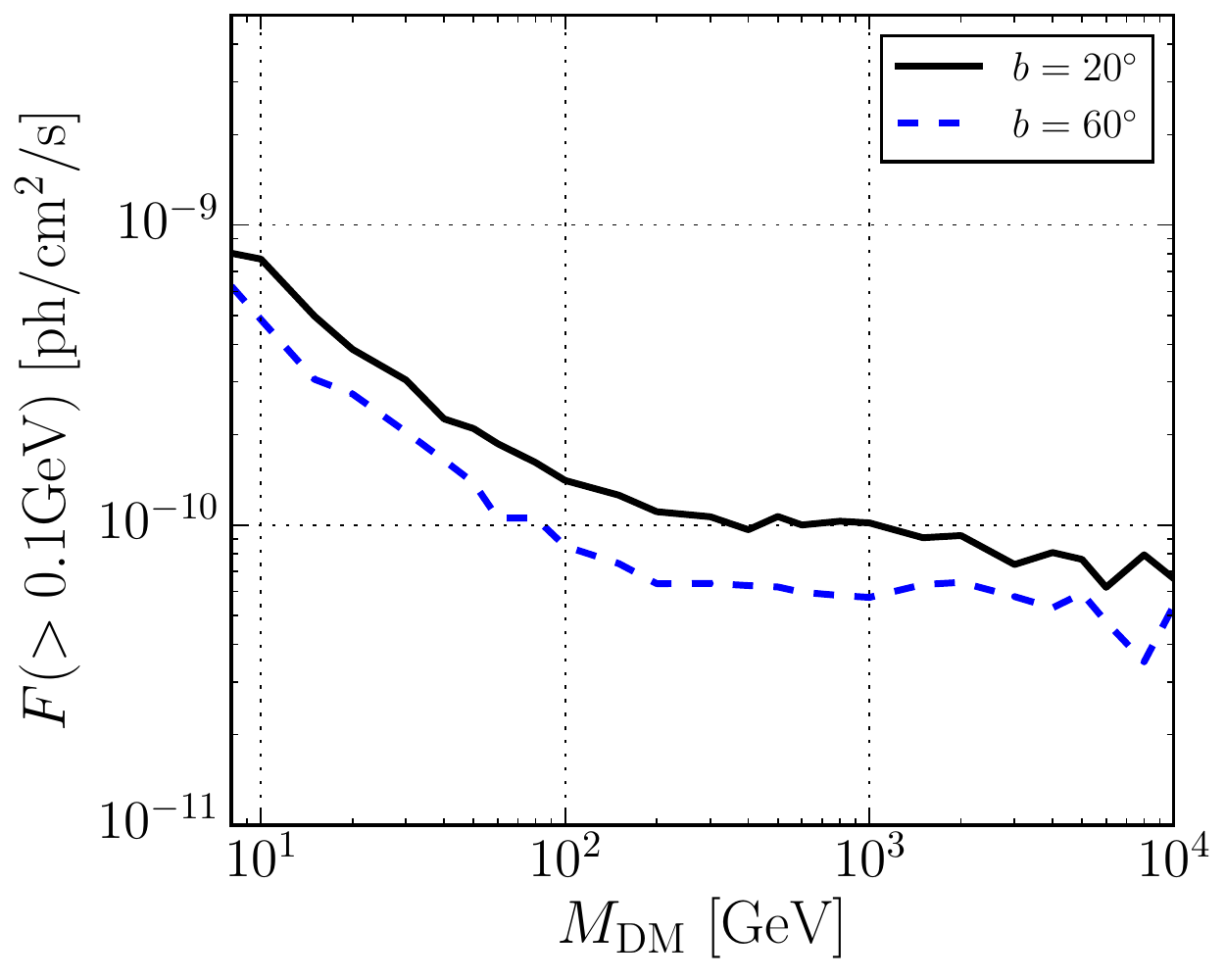}
\caption{Same as in figure~\ref{fig:sens3FGLbb} but for $\tau^+\tau^-$ annihilation channel.}
\label{fig:sens3FGLtau} 
\end{figure}

In the left panel of figure~\ref{fig:sens3FGLbb} (figure~\ref{fig:sens3FGLtau}) we show the sensitivity flux for $b\bar{b}$ 
($\tau^+\tau^-$) for a selection of DM masses,
 as a function of the Galactic latitude $b$. In the right panel, we report the same information but fixing two  
 latitude values.
For each DM mass, fluxes larger than a specific curve would be detected with TS $>$ 25. That means that a SH made of DM 
 particles of given mass could be resolved by the \Fermi-LAT (in its 3FGL configuration) 
  if the emitted flux above 0.1 GeV is above that threshold. 
For both annihilation channels, the flux sensitivity threshold is a mild decreasing function of the latitude.
As one moves away from the Galactic plane (i.e.~towards high latitudes) the intensity of the IEM decreases 
and therefore the detection of a fainter halo is easier because of the lower background. Additionally, the dependence 
of the flux sensitivity threshold on the DM mass is also peculiar: going from 8 GeV to about 300 GeV 
the sensitivity flux threshold decreases significantly (by a factor of  $\sim10$), while for DM masses 
larger than a few hundreds GeV the sensitivity flux decreases only slightly, and settles to values $\sim 10^{-10}$ 
ph/cm$^2$/s.
On the one hand, for small DM masses the flux sensitivity is larger because the slope of the DM gamma-ray 
spectrum is softer than  for heavier DM masses (cf.~figure~\ref{fig:dmspectrafit}) and, as explained above, 
the LAT detects smaller fluxes for sources with harder spectra at  $E>0.1$ GeV ~\cite{Collaboration:2010gqa}. 
Moreover, for DM masses above $\mathcal{O}$(100) GeV the peak of the energy 
spectrum is at energies where the LAT point spread function (PSF) becomes smaller and the acceptance larger. 
For example, the peak of the energy spectrum for annihilation into $b\bar{b}$ and DM mass of 10 GeV is at  
 $\sim400$ MeV where the PSF is 2$^{\circ}$ and the acceptance is about 2.25 m$^2$sr. On the other hand, for a 
candidate with a DM mass of 300 GeV, the gamma-ray energy spectrum peaks at  $\sim10$ GeV, where the PSF
 is 0.2$^\circ$ and the acceptance is 2.50 m$^2$sr.
The smaller size of the PSF and the larger acceptance explain the order of magnitude 
of difference in the sensitivity flux threshold between these two cases. 
Finally, for DM masses larger than 300 GeV the flux sensitivity decreases only mildly,
because the PSF and the acceptance at the position of the gamma-ray energy spectrum 
peak are worse. In this case, only the shape of the energy spectrum matters for the detection 
and all the considered mass candidates have similar spectral energy distributions.

\subsection{Sensitivity to dark matter fluxes for the 2FHL catalog setup}
\begin{figure}
	\centering
	\includegraphics[width=0.49\columnwidth]{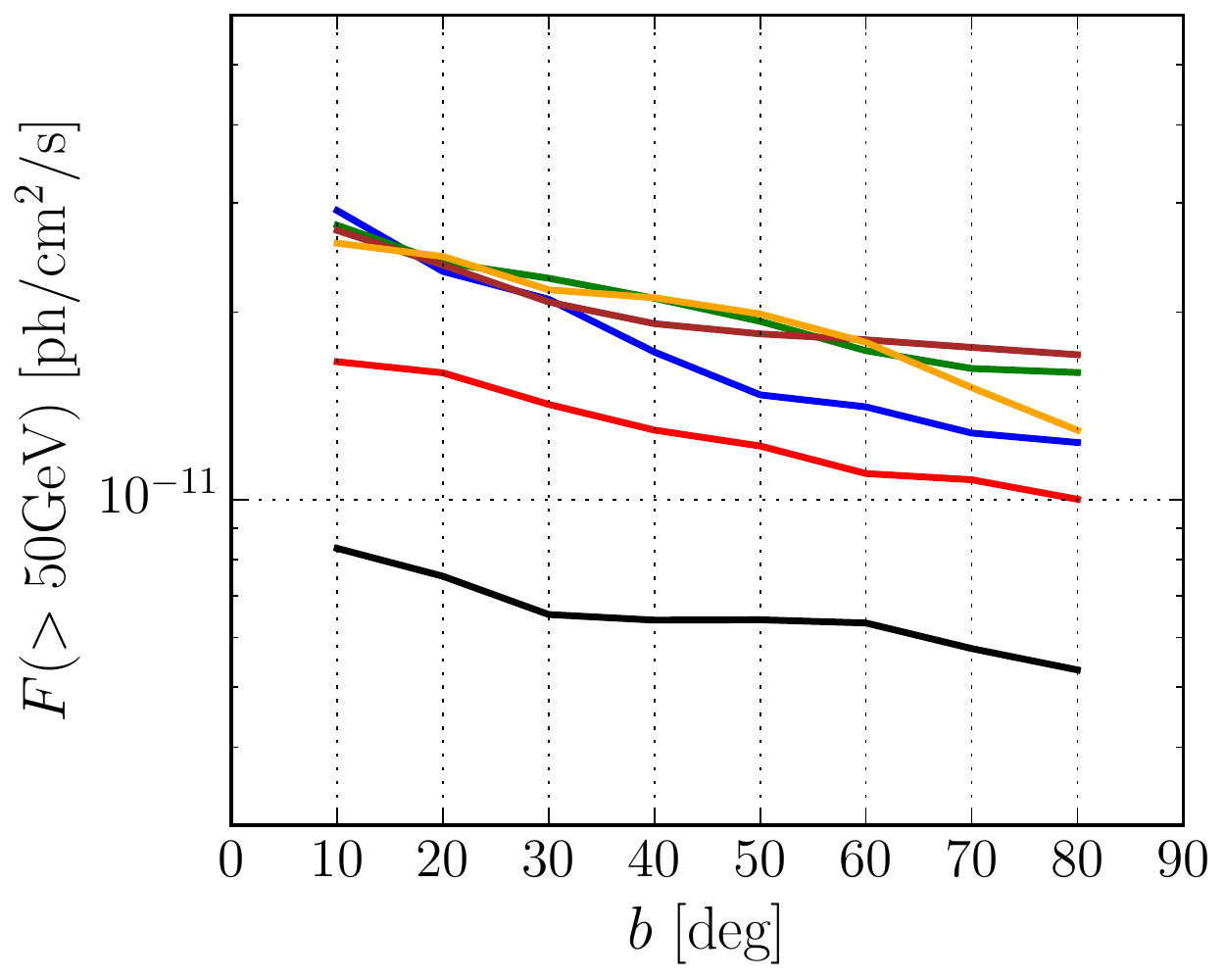}
	\includegraphics[width=0.49\columnwidth]{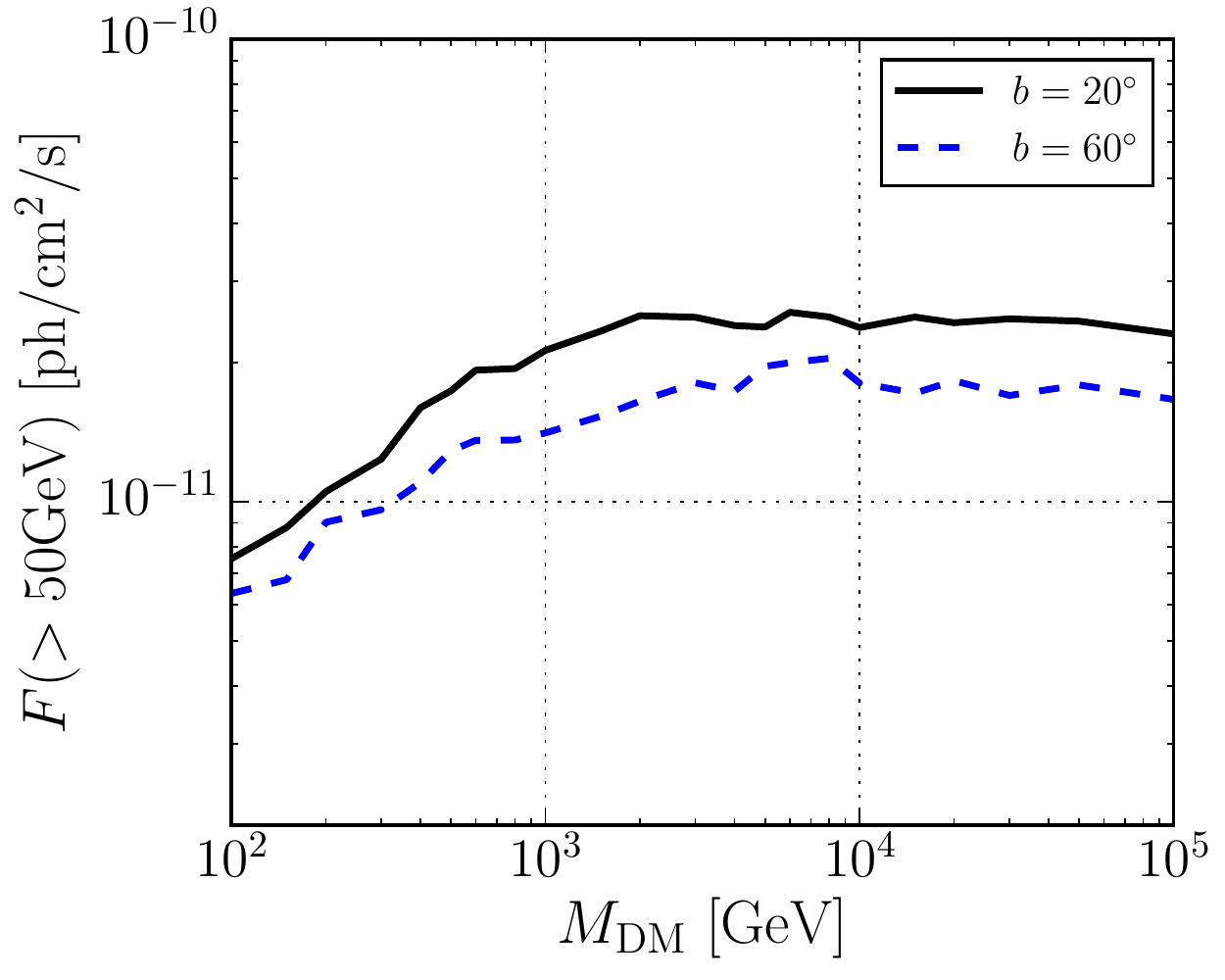}
\caption{Flux sensitivity threshold of \Fermi-LAT 2FHL to DM annihilation spectra for $b \bar{b}$ 
	annihilation channel.
	\emph{Left panel}: Flux sensitivity threshold as a function of position (latitude) of the SH for, from bottom to top, 
	$M_{\rm{DM}}=$ 100 (black), 400 (red), 1000 (blue), 4000 (green), 8000 (brown), 20000 GeV
	(orange).
	\emph{Right panel}: Flux sensitivity threshold as a function of DM mass for $b=20^{\circ}$ and $60^{\circ}$ of the SH.} 
\label{fig:sens2FHLbb} 
\end{figure}

\begin{figure}
	\centering
	\includegraphics[width=0.49\columnwidth]{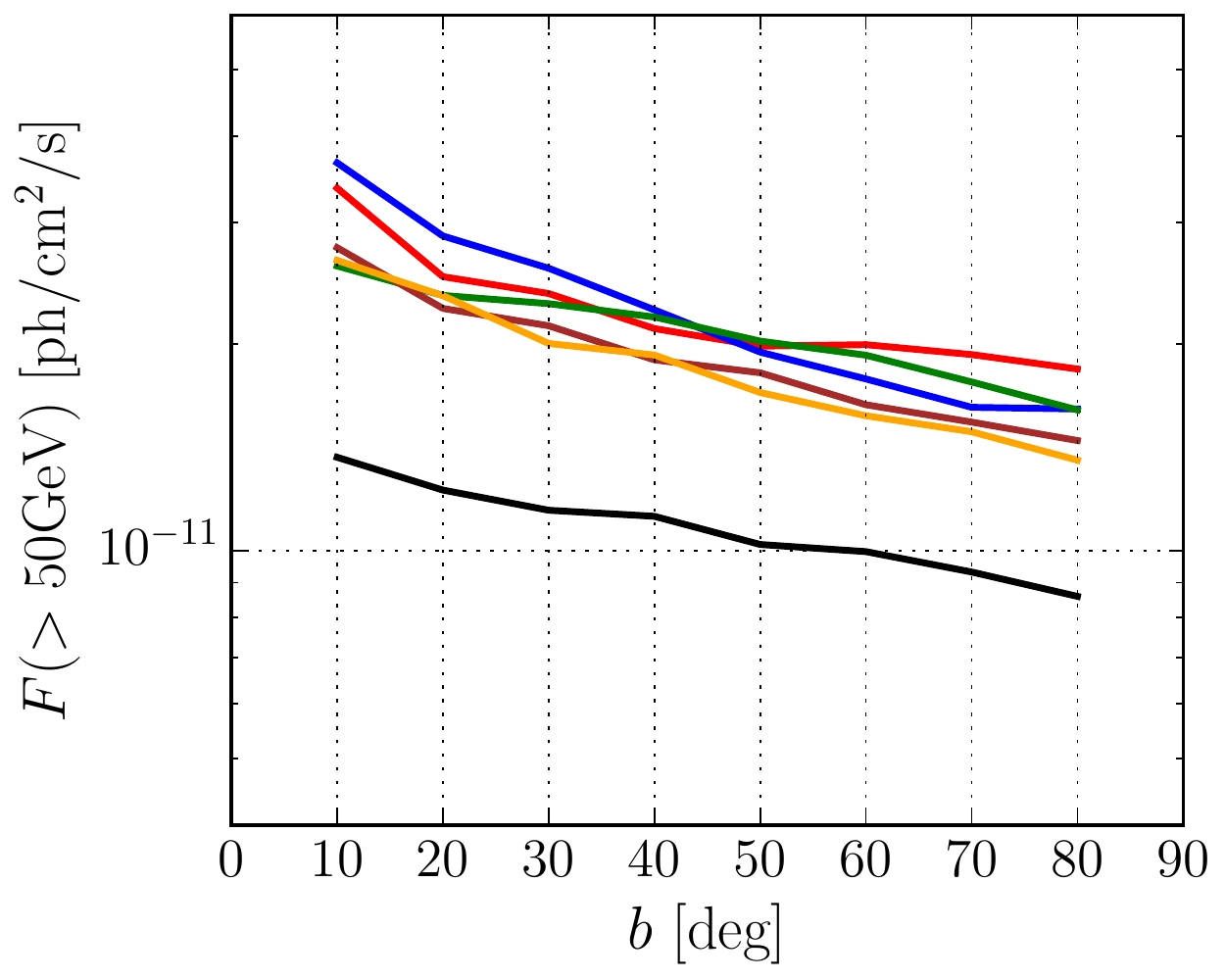}
	\includegraphics[width=0.49\columnwidth]{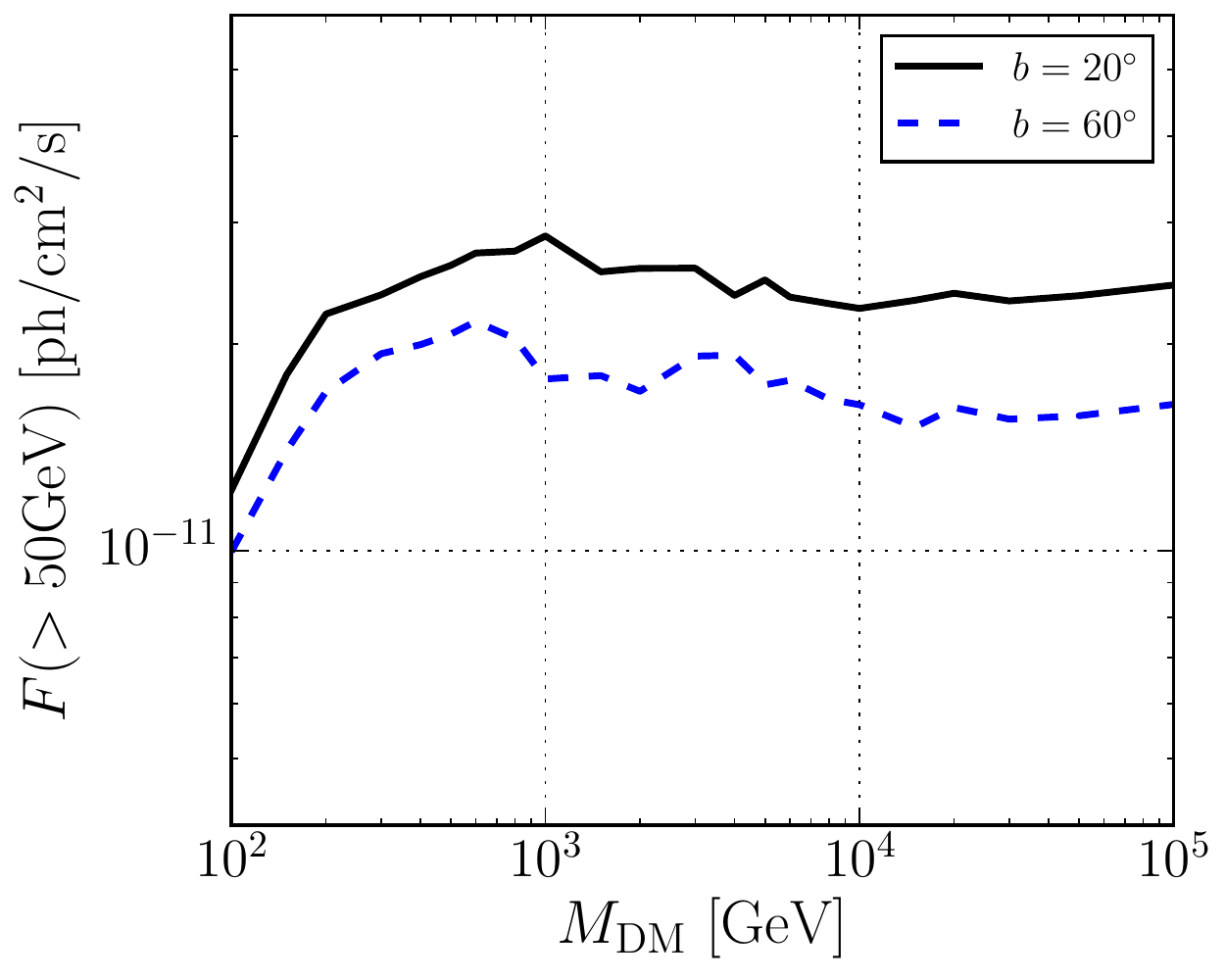}
\caption{Same as in figure~\ref{fig:sens2FHLbb} but for $\tau^+\tau^-$ annihilation channel.}
\label{fig:sens2FHLtau} 
\end{figure}

In this section we report the results for the flux sensitivity for the 2FHL catalog setup.
We have considered 80 months of data (from August 2008 to April 2015) and the energy 
range $50 - 2000$ GeV divided into 5 logarithmic energy bins.
The Pass 8 SOURCE class of data has been used with an ROI centered around each 
DM SH with a radius of 10$^{\circ}$ and a spatial binning $0.1\times 0.1$ deg$^2$.
We vary the DM mass between 100 GeV and 100 TeV, since DM masses smaller than 
100 GeV  have the most of the gamma-ray spectrum below 50 GeV. We use
 the {\tt gll\_iem\_v06.fits} and {\tt iso\_P8R2\_SOURCE\_V6\_v06.txt} templates.
In figure~\ref{fig:sens2FHLbb} (for $b\bar{b}$) and~\ref{fig:sens2FHLtau} (for $\tau^+\tau^-$) 
we report the flux sensitivity 
threshold as a function of latitude for a selection of DM masses (left panels), and  as a function of $M_{\rm{DM}}$ for fixed 
 $b=60^{\circ}$ and $20^{\circ}$ (right panels).
In the case of the 2FHL, the sensitivity profile shows an opposite trend with respect to the 
3FGL case, since it increases with  DM mass, reaching a plateau for $M_{\rm{DM}} \sim 1$ TeV. 
The flux sensitivity does not change for  DM masses $M_{\rm{DM}} > 1$ TeV. A DM SH made 
of TeV mass particles has the same chance to be detected by the \Fermi-LAT as a $\mathcal{O}$(10) TeV DM mass SH. 
The sensitivity flux threshold grows from 100 GeV to 1 TeV because the energy threshold for the 
2FHL is 50 GeV. In this energy range and for the considered DM masses, the gamma-ray spectrum 
has a very soft shape with peak  at $E<50$ GeV. Then, for $M_{\rm{DM}}>1$ TeV the peak falls
 inside the 2FHL energy range, the sensitivity flattens  and reaches a plateau. For $M_{\rm{DM}}>1$ 
 TeV the sensitivity threshold remains constant because the shape of the energy spectrum for these mass candidates is quite similar.

\section{Detectability of dark matter sub-halos}
\label{sec:results}
In this section we report our results for the detectability of DM SHs. We  give our predictions in terms of
 i) the number of detectable SHs in the 3FGL and 2FHL catalogs setups, ii) bounds on the DM 
 annihilation cross section and iii) the source count distribution of DM SHs, compared to the one of blazars.

\subsection{Number of detectable sub-halos and limits on dark matter annihilation cross section}
The SHs that are detectable by the LAT are those with a flux above the \Fermi-LAT 
sensitivity flux threshold (for a specific catalog setup) and which could potentially be among the {\it unassociated} 
sources in the 3FGL and 2FHL catalogs. The reference \Fermi-LAT sensitivity fluxes have been discussed in the previous section.
While the number of detectable SHs relates to the brightest end of the SH luminosity
function, the faintest SHs remain below threshold and thus only contribute to the total SH source count 
distribution.

\begin{figure}
	\centering  
	\includegraphics[width=0.49\columnwidth]{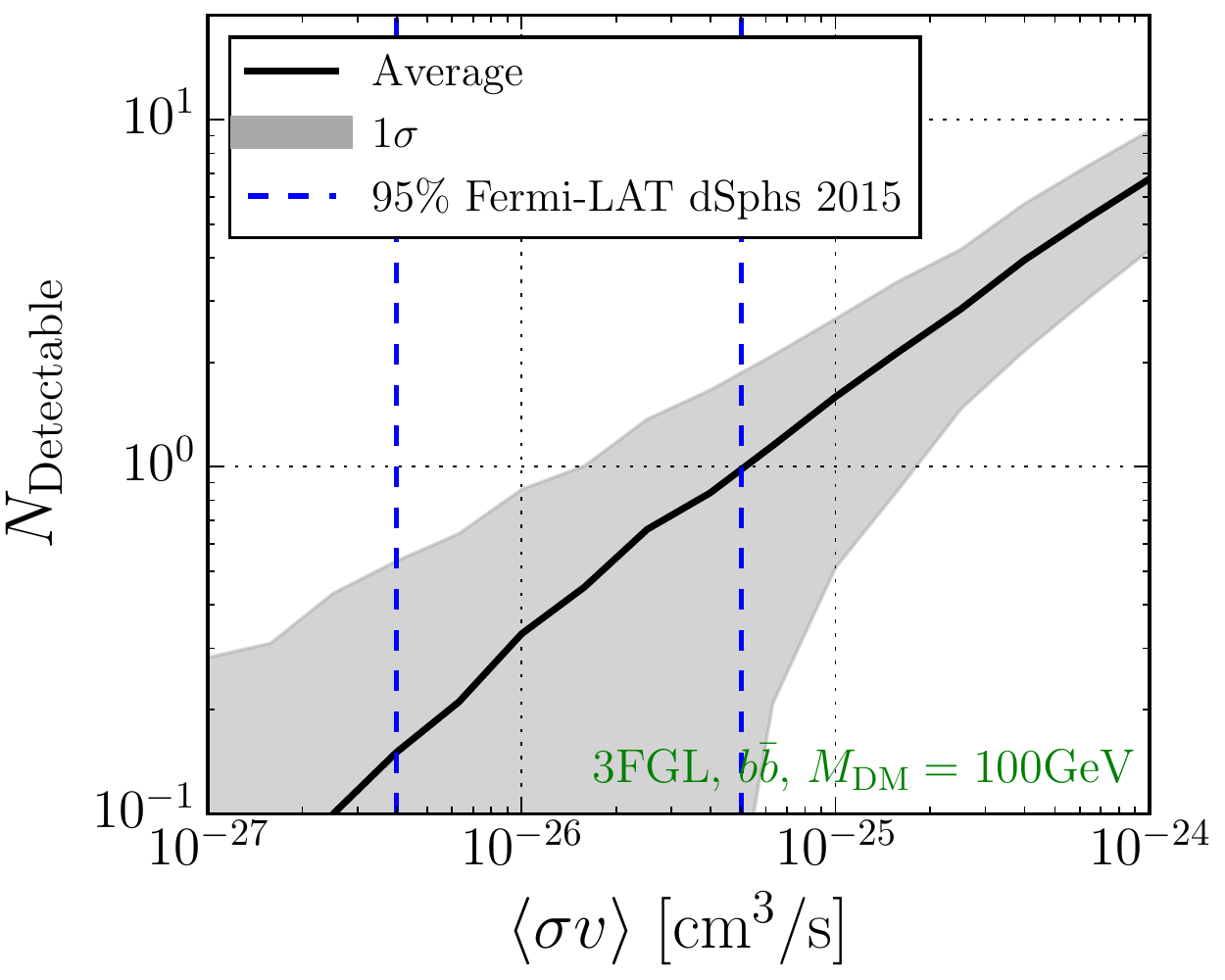}
	\includegraphics[width=0.49\columnwidth]{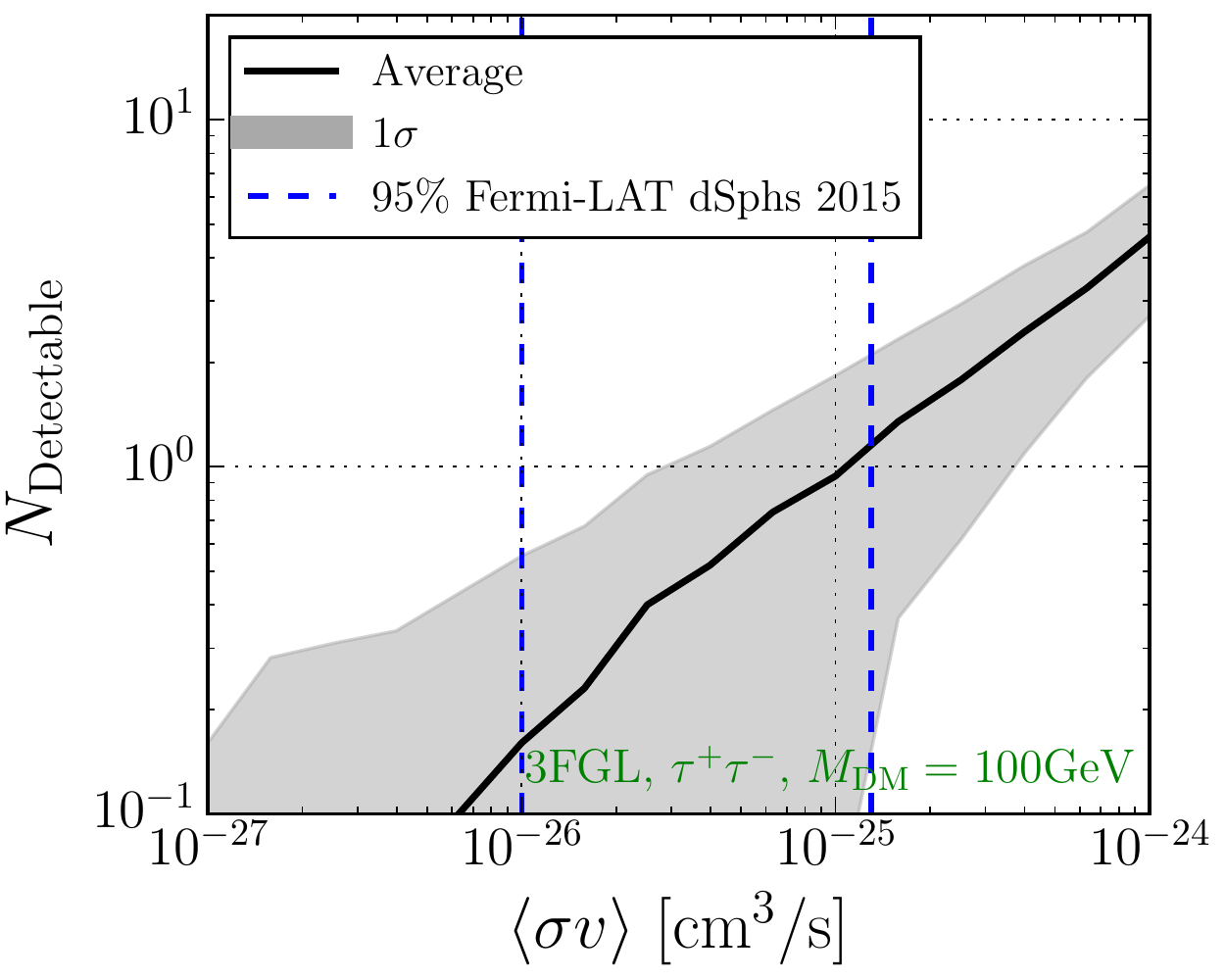}
	\includegraphics[width=0.49\columnwidth]{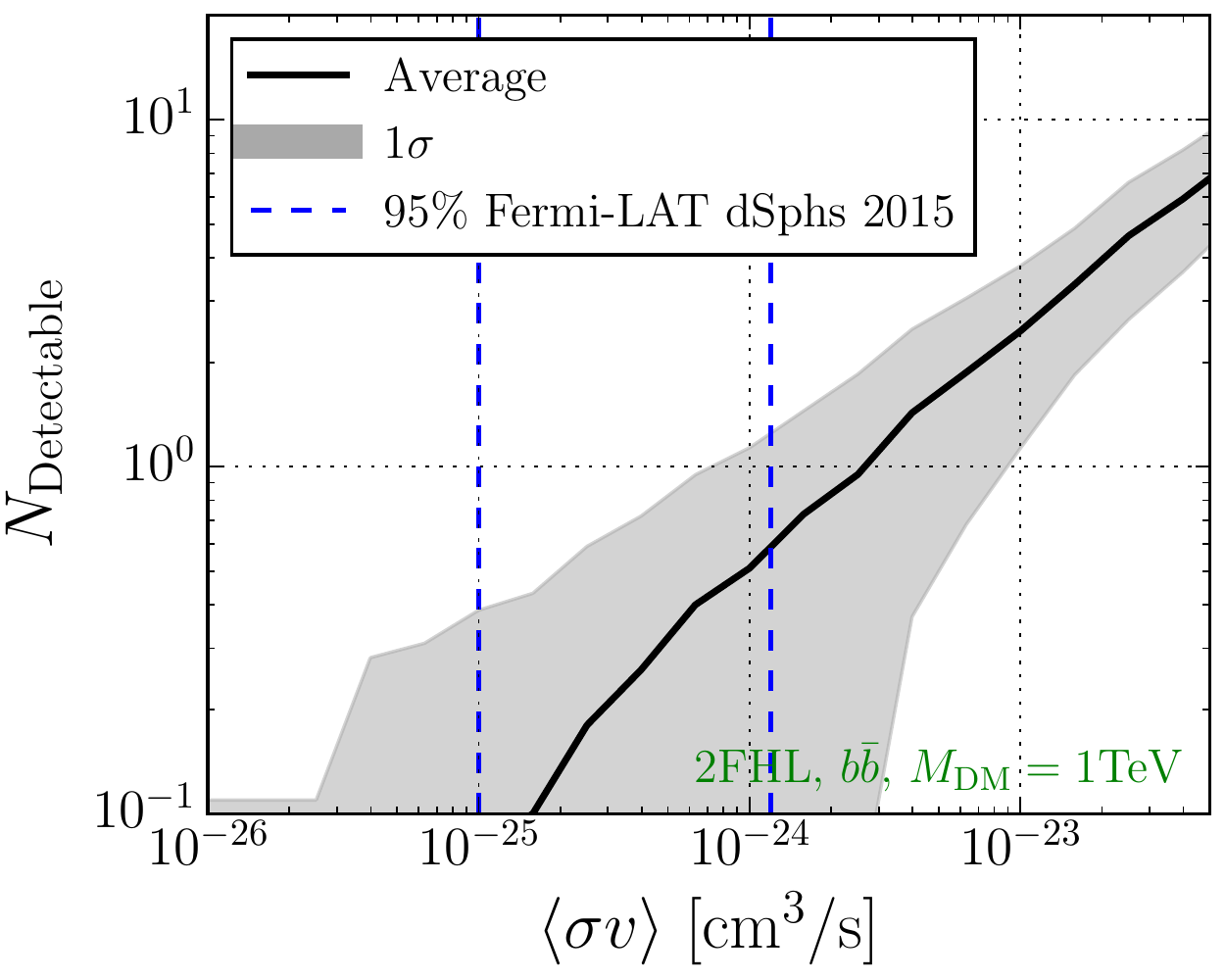}
\caption{
	Number of detectable SHs as a function of the annihilation cross section, $\sv$,
	for a fixed DM mass value.
	The black solid line represents the average over 100 Monte Carlo realizations of the
	SH population, while the grey band is the corresponding 1$\sigma$ uncertainty.
	The vertical dashed blue lines are the 95\% confidence level cross section upper limits from the 
	\Fermi-LAT dwarfs analysis~\cite{Ackermann:2015zua}.
	\emph{Top left panel}: $M_{\rm DM}$=100 GeV,  3FGL sensitivity, DM DM $\rightarrow b\bar{b}$. 
	\emph{Top right panel}: $M_{\rm DM}$=100 GeV, 3FGL sensitivity, DM DM $\rightarrow \tau^+ \tau^-$. 
	\emph{Bottom panel}: $M_{\rm DM}$=1 TeV,  2FHL sensitivity,  DM DM $\rightarrow b\bar{b}$.}
\label{fig:Ndetectedthree} 
\end{figure}

In order to derive the number of detectable SHs, for all the SHs 
in our Monte Carlo realizations (see section~\ref{sec:model})
we compute the gamma-ray flux above a given energy according to eq.~(\ref{eq:phflux}) 
and assuming an Einasto DM density profile in the SHs. We then compare the predicted 
flux with the sensitivity flux threshold, depending on the latitude of the individual SH, both for the
 3FGL and 2FHL setups, as derived in section~\ref{sec:sens}. A SH is defined as detectable 
 if the predicted gamma-ray flux is larger than the sensitivity flux threshold at the SH position. 

In figure~\ref{fig:Ndetectedthree}, we show the dependence of the number of detectable SHs 
on the annihilation cross section for different choices of the DM mass, DM annihilation channel and
catalog sensitivity. 
The number of detectable SHs increases with the $\sv$ almost linearly. 
If we consider the 95\% confidence level upper limits on $\sv$ from the \Fermi-LAT
analysis of dwarf galaxies~\cite{Ackermann:2015zua}, about one SH could be present in the \Fermi-LAT catalogs.

In particular, we checked that, fixing the annihilation cross section to the one constrained by the dwarfs analysis~\cite{Ackermann:2015zua} and assuming the 3FGL catalog sensitivity, the number of detectable SHs 
is only mildly dependent on the DM mass if we assume a $b\bar{b}$ annihilation channel, while it 
decreases more rapidly as a function of the DM mass for DM DM $\rightarrow \tau^+ \tau^-$.
In the case of the 2FHL, instead, the number of detectable SHs slightly increases with DM mass.
However,  in all cases we deal with very small numbers of detectable SHs, i.e~$\lesssim$ $\mathcal{O}$(1).
The number of detectable SHs that might already be among the unassociated sources of the 3FGL catalog
turns out to be $0.9\pm 0.8$ for $M_{\rm{DM}}=8$ GeV. For the 2FHL, $N_{\rm{Detectable}}$ is even smaller: 
$0.0\pm0.2$ for $M_{\rm{DM}}=10$ TeV. 
These are very small numbers compared to the amount of unassociated sources in the 3FGL (1062) and 2FHL (48) catalogs, 
and are compatible with the fact that no emission from the direction of known dwarf galaxies has been
observed yet.

Such small numbers of detectable SHs are  lower than what found in the literature, mostly because  
here we fully model the sensitivity of the \Fermi-LAT to DM SHs, as explained in section~\ref{sec:sens}.
We checked that using a fixed energy flux detection threshold -- as given by the energy flux integrated above 
1 GeV\footnote{We use this quantity in order to reduce the bias between the source flux and 
the photon index, see \cite{Collaboration:2010gqa}.} and equal to the minimum flux of sources (in the 3FGL), 
$4.0\cdot 10^{-13}$ erg/cm$^2$/s -- we get twice more detectable SHs. On the other hand, 
using the energy flux that gives the peak of the energy flux distribution, namely 
$1.35\cdot 10^{-12}$ erg/cm$^2$/s, leads to $20\%$ less detectable SHs for 
$M_{\rm{DM}}=100$ GeV with respect to the former optimistic threshold.

\begin{figure}
	\centering  
	\includegraphics[width=0.49\columnwidth]{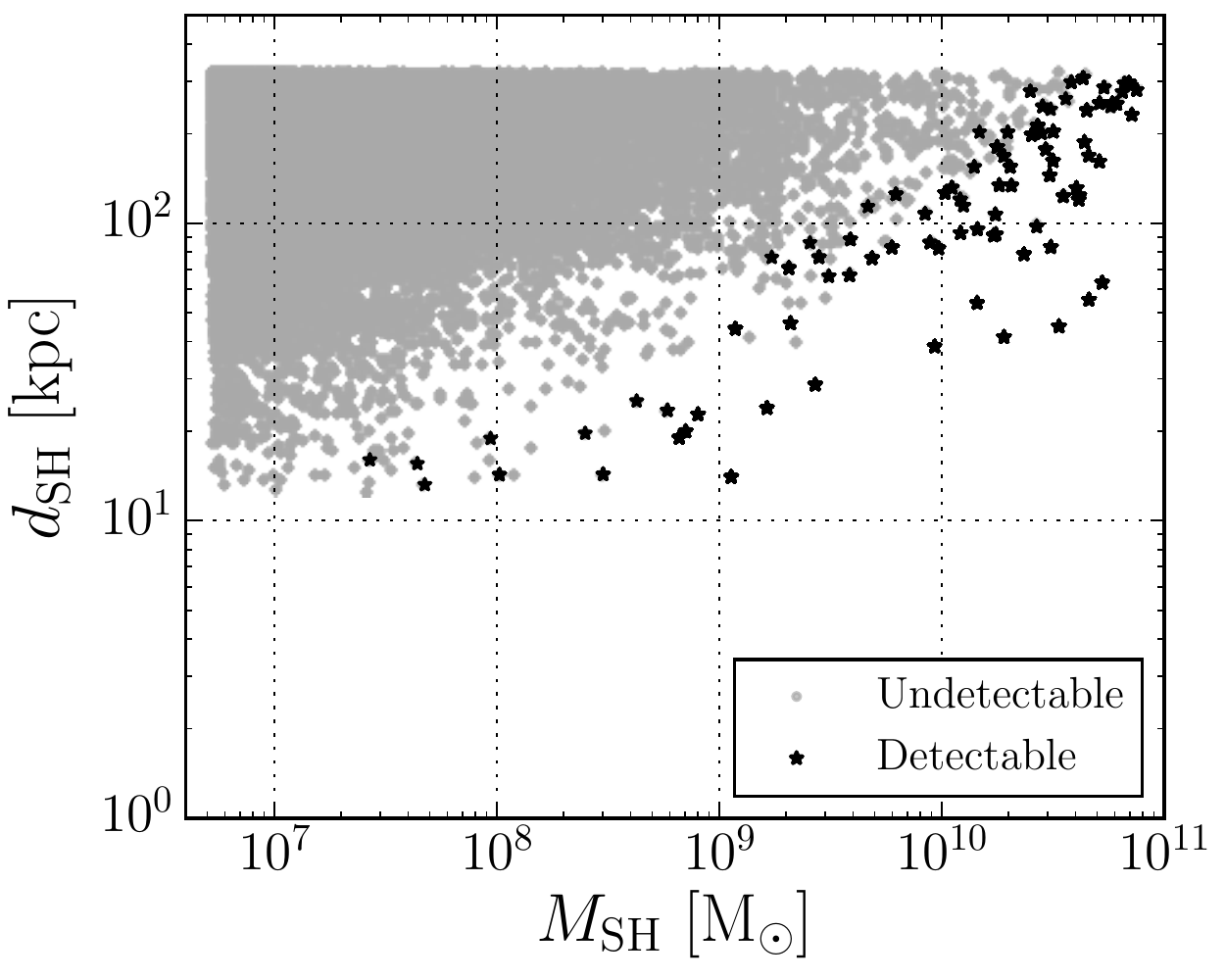}
      \includegraphics[width=0.49\columnwidth]{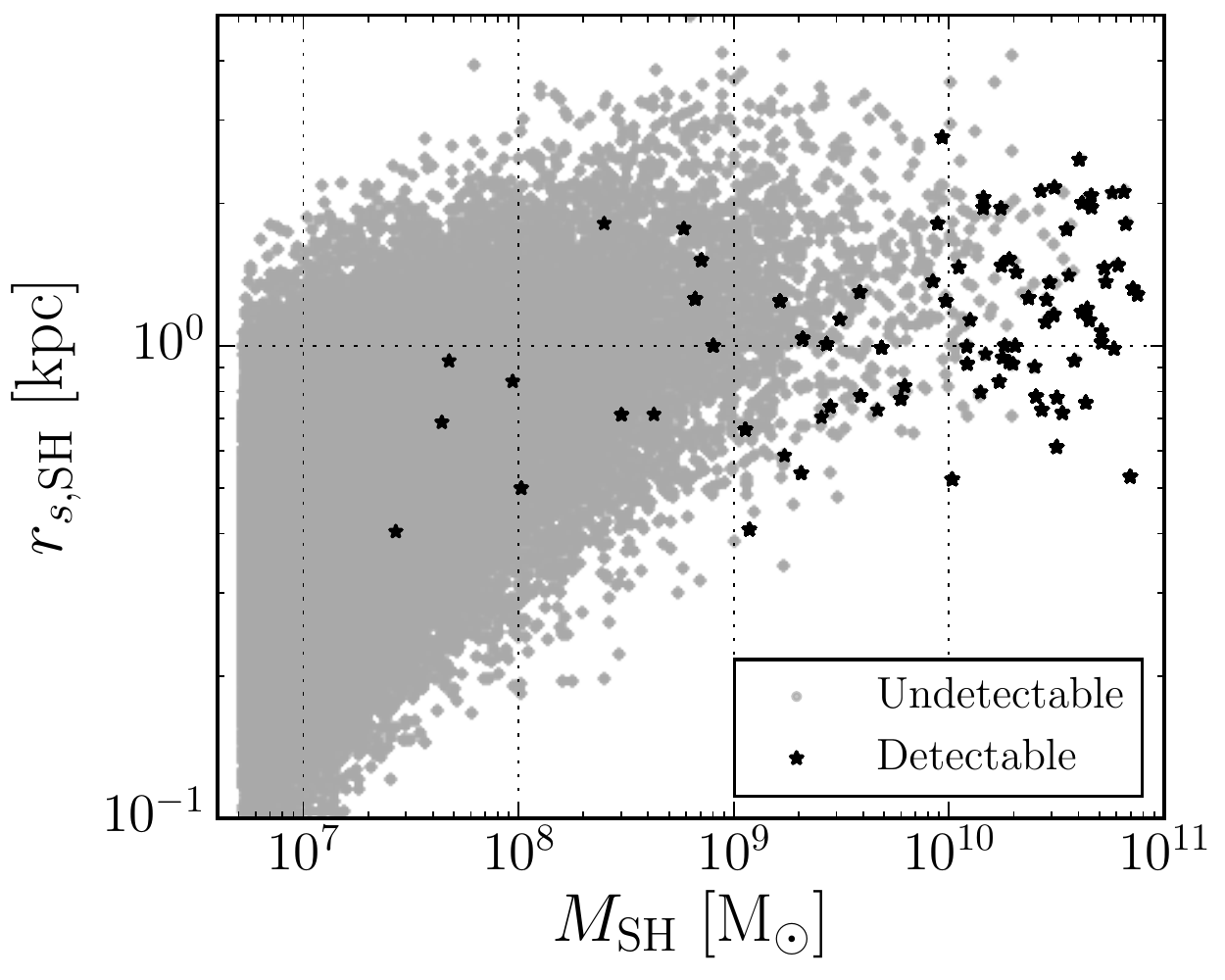}
\caption{\emph{Left panel}: SH distance from the observer, $d_{\rm SH}$, as a function of the SH mass, 
	$M_{\rm SH}$, for the detectable SHs (black stars) and the SHs below threshold (grey points).
	\emph{Right panel}: Scale radius $r_s$ as a function of $M_{\rm SH}$ for the detectable SHs 
	(black stars) and the SHs below threshold (grey points). 
	We adopt the 3FGL sensitivity and DM annihilation into $b\bar{b}$.
	The results are shown for all 100 Monte Carlo realizations of the SH population.}
\label{fig:dmasssub} 
\end{figure}

\medskip
In figure~\ref{fig:dmasssub} we display the distance to the observer $d_{\rm SH}$  
vs the mass $M_{\rm SH}$ for all SHs in our 100 Monte Carlo realizations, and highlight
 the ones with a flux larger than the \Fermi-LAT 3FGL sensitivity flux.
We obtain few detectable SHs (depicted with black stars), with distances $d_{\rm SH}\in[10,300]$ 
kpc and $M_{\rm SH}> 2\cdot 10^7M_{\odot}$.
Contrary to what assumed in previous analyses (see e.g. \cite{Bertoni:2015mla}) we find that the
 detectable dark and luminous SHs may be more massive than $\sim 10^7$ \Msun. We wish to 
 stress that even if the minimum mass for SHs to host star formation is about $10^{7.5}$ \Msun, 
 dark SHs (i.e. without stars) are realized in the simulation up to masses $\sim 10^9$ \Msun, and 
 hence coexist together with luminous SHs in the mass range $10^{7.5} h^{-1} -10^9 h^{-1}$\Msun~\cite{Zhu:2015jwa}.
 Larger mass SHs are instead much more likely to have a stellar counterpart and therefore 
 to be detected in the optical wavelength as dwarf galaxies.
We also show in  figure~\ref{fig:dmasssub} the value of the scale radius $r_s$ of each SH vs the SH mass. 
The smallest $r_s$ values correspond to undetectable SHs, independently of $M_{\rm SH}$. 
Detectable SHs can have $r_s$ ranging from 
0.4 kpc to 3 kpc, regardless the value of the SH mass.

\begin{figure}
	\centering
	\includegraphics[width=0.49\columnwidth]{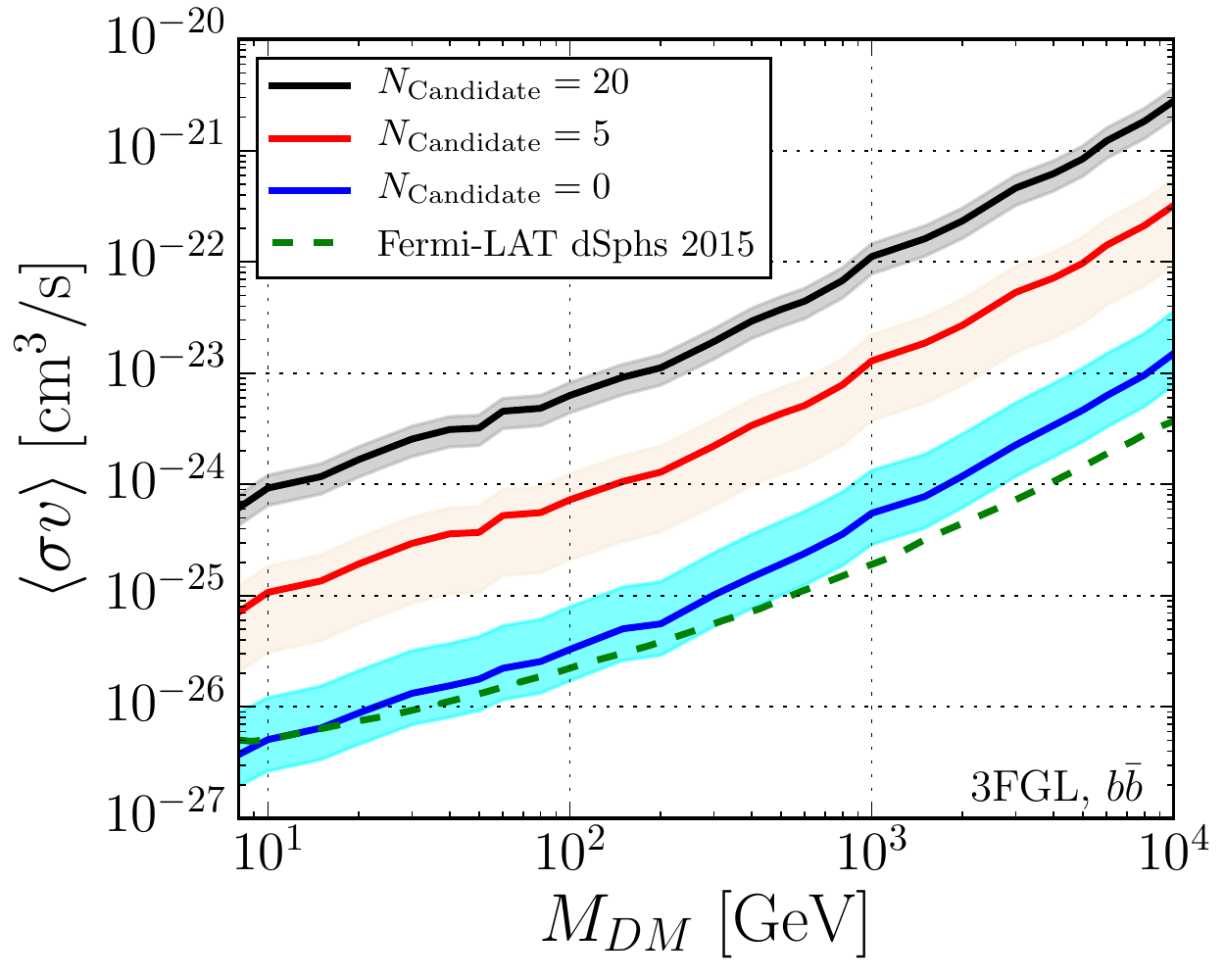}
	\includegraphics[width=0.49\columnwidth]{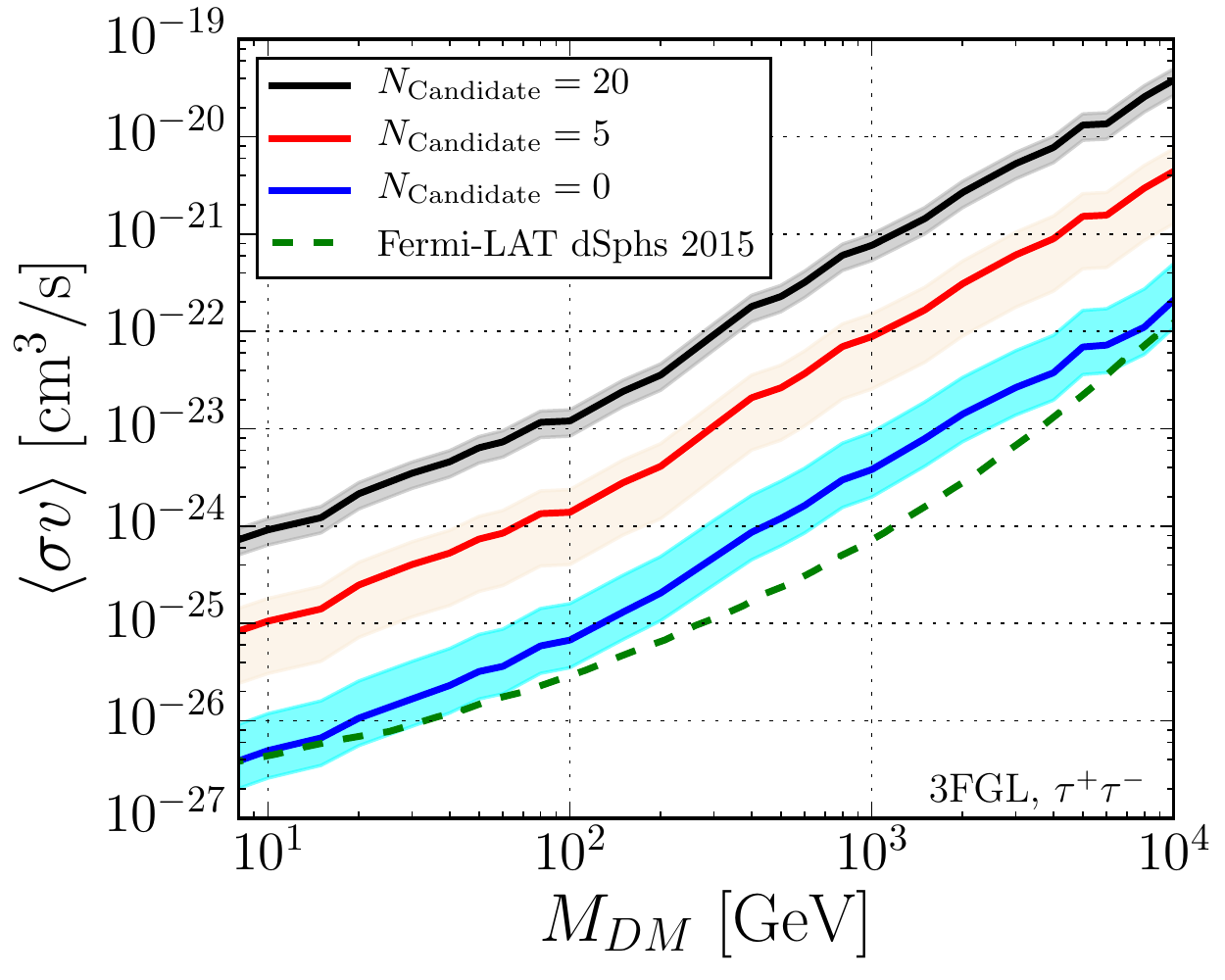}
	\includegraphics[width=0.49\columnwidth]{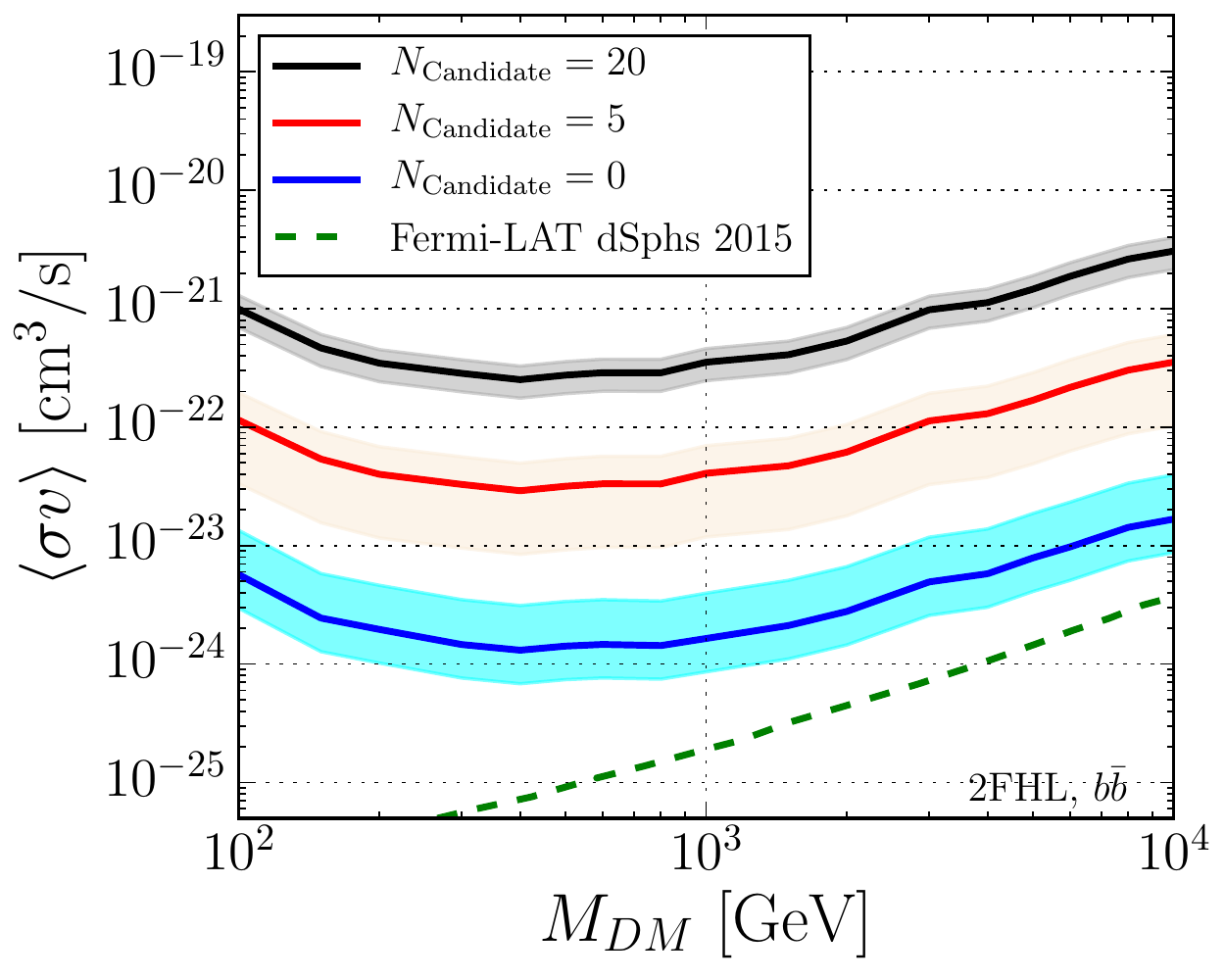}
\caption{Upper limits on $\sv$ derived assuming 20 (grey),  5 (red) and 0 (cyan) SHs candidates in \Fermi-LAT catalogs, 
	together with the bounds from the dwarf galaxies \Fermi-LAT analysis~\cite{Ackermann:2015zua}. 
	\emph{Top left (right) panel}: Annihilation into $b\bar{b}$ ($\tau^+ \tau^-$) and 3FGL catalog setup.
	\emph{Bottom panel}: Annihilation into $b\bar{b}$ and 2FHL catalog setup.
	}
\label{fig:sigmav3fgl} 
\end{figure}

\medskip
The small (or even null) number of detectable DM SH candidates among the \Fermi-LAT 
unassociated sources allows us to set upper limits on the DM annihilation cross section $\sv$.
For each DM mass, we define as upper limit the value of $\sv$ for which the number of 
detectable SH is smaller than a given number $N_{\rm Candidate}$ of DM SH candidates.  
Should $N_{\rm Candidate}$ be zero, the most stringent constraints on $\sv$ would be inferred. 
However, the number of unassociated sources in the two catalogs is 
not zero, and we do expect some DM SHs among them. Indeed, the case in 
which $N_{\rm Candidate}=N_{\rm Unassociated}$ would give the most conservative upper limits, 
not accounting for the fact that many unassociated sources 
are very likely going to be identified as standard astrophysical objects. 

In the following, we will show upper limits on $\sv$ assuming  $N_{\rm Candidate}$ = 0, 5 and 20.
We consider the number $N_{\rm Candidate}$ of brightest SHs (in terms of \Jf) for all the 100 Monte 
Carlo realizations, and we define the upper limit 
on $\sv$ as the maximum value of $\sv$ for which the SHs fluxes are equal to the sensitivity flux 
thresholds -- for a given catalog  --
at the corresponding SHs position.
We depict in figure~\ref{fig:sigmav3fgl} the upper bounds on the $\sv$, assuming the possible 
detection of 0 (cyan), 5 (red) and 20 (grey) SHs, for the 3FGL catalog setup (upper panels) and DM
annihilation channel into $b\bar{b}$ (left panel) or $\tau^+\tau^-$ (right panel), and for the 2FHL catalog setup
for $b\bar{b}$ annihilation channel (bottom panel).  
The bounds for the detection of $N_{\rm Candidate}$ = 5 and 20 result weaker than those 
derived with the Pass 8 analysis of dwarf galaxies~\cite{Ackermann:2015zua}. On the contrary, 
the limits derived assuming $N_{\rm Candidate}$ = 0 are very tight
and competitive with limits from dwarfs galaxies. 
The reason is that the brightest SH in all realizations has a very high flux. 
For example, the gamma-ray flux of the brightest SH with DM
mass of 100 GeV and with thermal cross section is on average $1.6\times10^{-9}$ ph/cm$^2$/s, 
thus above the sensitivity threshold at  b=$30^{\circ}$  ($\sim 7\times10^{-10}$ ph/cm$^2$/s, cf.~figure~\ref{fig:sens3FGLbb}).

The dependence of the cross section upper limits on the DM mass can be understood as follows:
The annihilation cross section is derived from eq.~(\ref{eq:phflux}) as $\sv \approx 
(\Phi 4 \pi M^2_{\rm{DM}})/(\mathcal{J} \mathcal{I})$, and, taking into account only the quantities 
dependent on the DM mass, $\sv \propto ( \Phi M^2_{\rm{DM}})/(\mathcal{I})$. The integrated 
gamma-ray energy spectrum from DM annihilation for $b\bar{b}$ channel is $\mathcal{I} 
\propto M_{\rm{DM}}^{0.4}$. On the other hand, the sensitivity flux goes as $\Phi \propto M_{\rm{DM}}
^{-0.8}$ for $M_{\rm{DM}}<100$ GeV and $\Phi \propto M_{\rm{DM}}^{-0.4}$ for $M_{\rm{DM}}>100$ 
GeV (see figure~\ref{fig:sens3FGLbb}). Therefore, $\sv \propto M_{\rm{DM}}^{0.8}$ for $M_{\rm{DM}}<
100$ GeV and $\sv \propto M_{\rm{DM}}^{1.2}$ for $M_{\rm{DM}}>100$ GeV, as shown in figure~\ref{fig:sigmav3fgl}.

In full analogy, it is possible to explain the trend of the upper limits in the case of the 2FHL.
In this case, for $M_{\rm{DM}}\in[100,500]$ GeV, $\mathcal{I} \propto M_{\rm{DM}}^{3.0}$ and 
$\Phi \propto M_{\rm{DM}}^{0.6}$ so that the annihilation cross section decreases as  $\sv \propto 
M^{-0.4}$. On the other hand, for $M_{\rm{DM}}>500$ GeV the flux sensitivity flattens (see 
figure~\ref{fig:sens2FHLtau}) and $\mathcal{I} \propto M_{\rm{DM}}^{1.0}$, so that  roughly  
$\sv \propto M^{1.0}$ as observed in figure~\ref{fig:sigmav3fgl} (bottom panel).

With a larger number of DM SHs candidates, the bounds reported in figure~\ref{fig:sigmav3fgl} get looser 
and increase less steeply. This fact 
has important consequences: First of all, in the 3FGL catalog  there are about 1000 unassociated 
sources and decreasing this number -- even by a factor of 10 -- would not have a large impact on 
the upper limits inferred on $\sv$. On the other hand, in the 2FHL catalog  there are about 50 unassociated 
sources: reducing the number of unassociated sources in this  catalog  by a factor of two would improve 
the bounds on $\sv$ by a factor of almost 10. 

\medskip

Future gamma-ray experiments, such as CTA~\cite{Acharya:2013sxa} at TeV energies and 
new concept Compton-Pair Production Telescopes like Compair~\cite{Moiseev:2015lva} and e-ASTROGAM~\cite{DeAngelis:2016slk} 
at the MeV scale, will improve on the sensitivity to detect point sources and DM SHs. 
As mentioned in section~\ref{sec:model}, DM SHs are classified into dwarf galaxies (i.e.~luminous SHs) or
dark SHs according to the presence or absence of a stellar component. 
The lower (non-zero) stellar mass of the Hydro selected SHs is $1.4 \times 10^4 M_\odot$. 
We here estimate the probability to detect dwarf galaxies as DM SHs with a future gamma-ray instrument 
with a factor of 5 better sensitivity than the LAT above 100 MeV. This improvement could be achieved by 
e-ASTROGAM or Compair at MeV energies.  We consider DM annihilation into $b\bar{b}$ for DM mass of 100 GeV 
and thermal cross section. 
In the Hydro simulation, the fraction of luminous SHs
in the mass bins $M_{\rm SH} = [10^{6.7}-10^{7},10^{7}-10^{8},10^{8}-10^{9},10^{9}-10^{10},10^{10}-10^{11}$] 
is $N_{\rm SH}(M_* \geq 1.4 \times 10^4 M_\odot)/N_{\rm SH}$ = [0.000, 0.024, 0.409, 0.857, 1.000]. 
Running our analysis for the SHs detectability with an improved flux sensitivity, we find that the average number of 
detectable SHs in each bin of mass is [0.0, 0.24, 0.34, 0.57, 0.91].
Combining these two results together, we obtain that 2.1 SHs would be detectable on average and 1.5 out of these would be
dwarf galaxies. Therefore, given the input of the adopted simulation, a future gamma-ray experiment 
a factor of 5 more sensibile than the LAT has the power 
to detect a few SHs, with a probability of 75$\%$ to detect a dwarf galaxy.

\subsection{The Log\,N -- Log\,F relationship for dark matter sub-halos}
An important characterization of astrophysical source populations is given by the so-called ${\rm Log\,N}$ -- ${\rm Log\,F}$, or 
the source count distribution $N$ as a function of the integrated flux $F$, 
which can provide information also about the faintest end of the flux distribution for a  specific source population. 
\begin{figure}
	\centering
	\includegraphics[width=0.48\columnwidth]{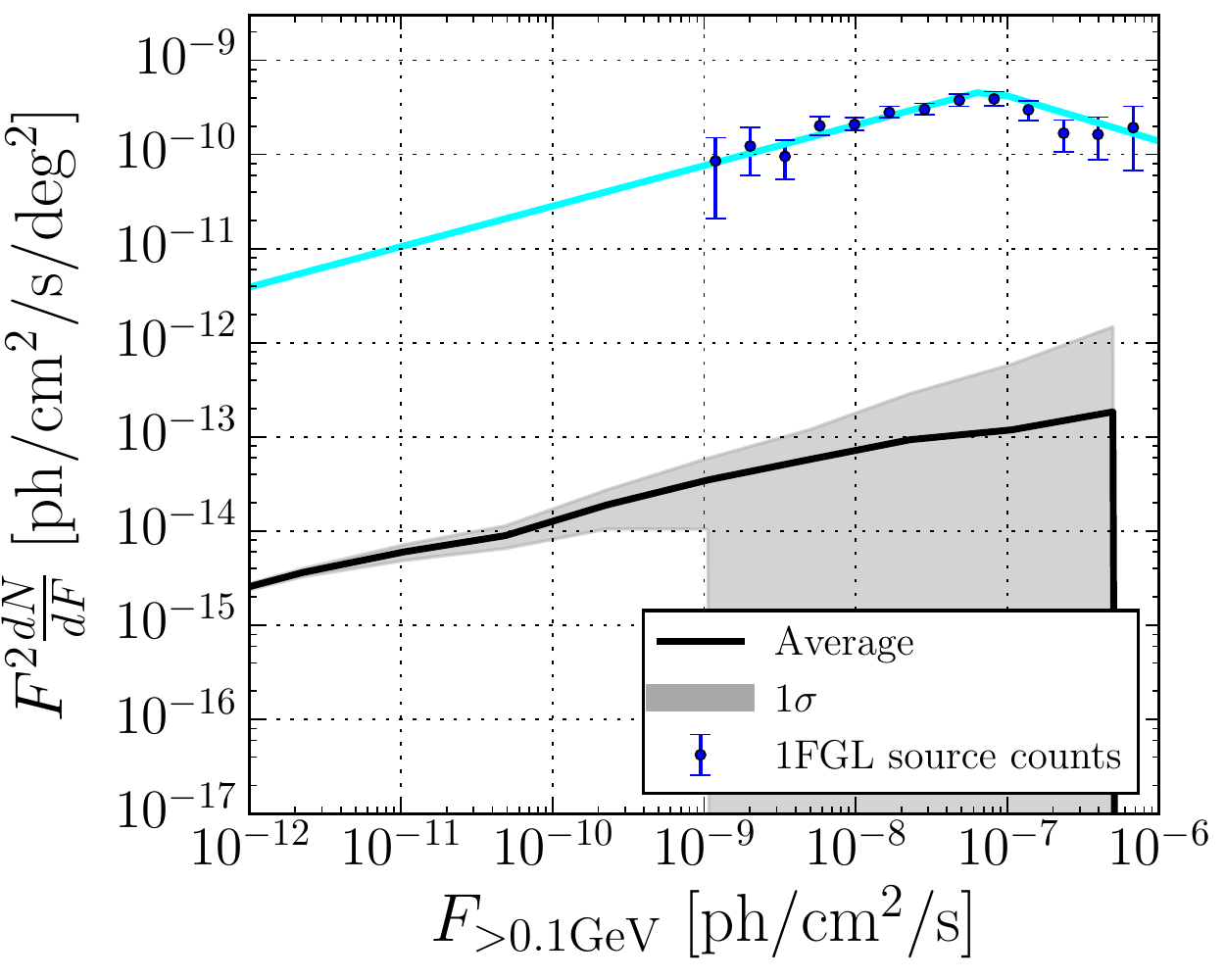}
	\includegraphics[width=0.49\columnwidth]{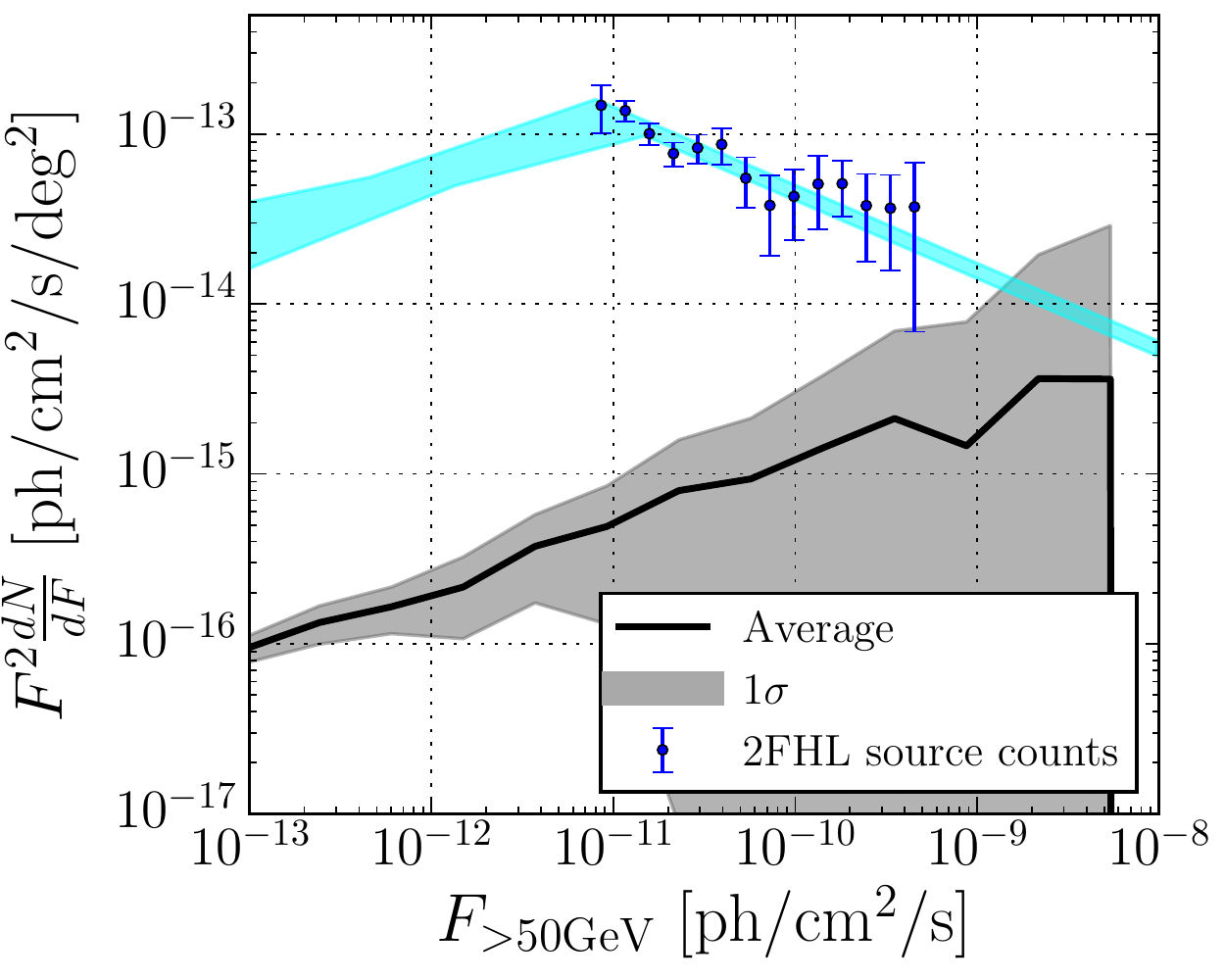}
	\caption{Source count distribution, or ${\rm Log\,N}$ -- ${\rm Log\,F}$, of all SHs in the mock Galactic SH 
	population for DM annihilation into $b\bar{b}$.
	\emph{Left panel}: 3FGL catalog setup, $M_{\rm DM}$=100 GeV and $\sv=10^{-25}$ cm$^3$/s. 
	The blue solid line represents the best-fit to the ${\rm Log\,N}$ -- ${\rm Log\,F}$ of the 
	blazars population in the 1FGL~\cite{Collaboration:2010gqa}. The black solid line is the average over 
	100 Monte Carlo realizations of the SH populations, while the grey band is the 
	corresponding 1$\sigma$ uncertainty.
	\emph{Right panel}: Same as in the left panel but with the 2FHL catalog setup
	and source count distribution of blazars as derived in ref.~\cite{TheFermi-LAT:2015ykq}.}
	\label{fig:lognlogsone} 
\end{figure}
For all the simulated SHs in the 100 Monte Carlo realizations of a Galactic SH population, 
we compute the photon flux $F$ as given by eq.~(\ref{eq:phflux}). 
We derive $dN/dF$ choosing a binning of the photon flux and considering for each $i$-th 
bin $\frac{dN}{dF}(F_i) = N_i/\Delta_i$, where $F_i$ is the center of the flux bin with a width 
$\Delta_i$ and $N_i$ is the number of SHs with a given flux in that bin.
For each flux bin we compute the mean and the standard deviation of $N_i$ over all Monte 
Carlo realizations, and we estimate the average and the $1\sigma$ dispersion for the $dN/dF$.
Finally, we compare this result with the same observable derived for AGN in the 1FGL~\cite{Collaboration:2010gqa} and 2FHL~\cite{TheFermi-LAT:2015ykq} catalogs.

In figure~\ref{fig:lognlogsone} we show the ${\rm Log\,N}$ -- ${\rm Log\,F}$ of all simulated DM SHs, with integrated 
flux above 0.1 GeV and 50 GeV respectively for the 3FGL (left panel) and 2FHL (right panel) catalog setups.
For comparison, we overlay the expected source count distribution from blazars in the
1FGL~\cite{Collaboration:2010gqa} and the recent estimate for high-energy blazars from the 2FHL~\cite{TheFermi-LAT:2015ykq}.
We consider annihilation into $b \bar{b}$ and DM mass of 100 GeV for the 3FGL and the 2FHL. 
The cross section is fixed to $\sv=10^{-25}$ cm$^3$/s.
The ${\rm Log\,N}$ -- ${\rm Log\,F}$ of DM SHs shows a sharp cutoff at high fluxes, that 
corresponds to few very bright SHs -- in the case
of the chosen annihilation cross section this is at about $5 \times 10^{-7}$
($5 \times 10^{-9}$) ph/cm$^2$/s for integrated fluxes above 0.1 (50) GeV.
The numerous faint and undetectable SHs populate the ${\rm Log\,N}$ -- ${\rm Log\,F}$ at low fluxes.
Regardless of the choice of the integration energy threshold, the SHs source count is strongly  
subdominant with respect to the observed flux distribution of AGN in both the 3FGL and 2FHL catalogs.
This effect becomes stronger when considering lower values of $\sv$, 
which are consistent with current limits from dwarf galaxies.

\subsection{On the relevance of the smaller scales: $M_{\rm{SH}}>10^5$ \Msun}
As already mentioned in section~\ref{sec:DMmodelflux}, the hydrodynamic 
simulation studied in this work has a mass resolution of $5.4 \times 10^6$ \Msun.
Although we are mostly interested in analyzing the differences between the Hydro and DMO runs, 
 usually expected to be important for quite massive SHs, we anyhow investigate the effect
of lower-mass SHs. We proceed arguing for the Hydro case only. 

\begin{figure}
	\centering
	\includegraphics[width=0.55\columnwidth]{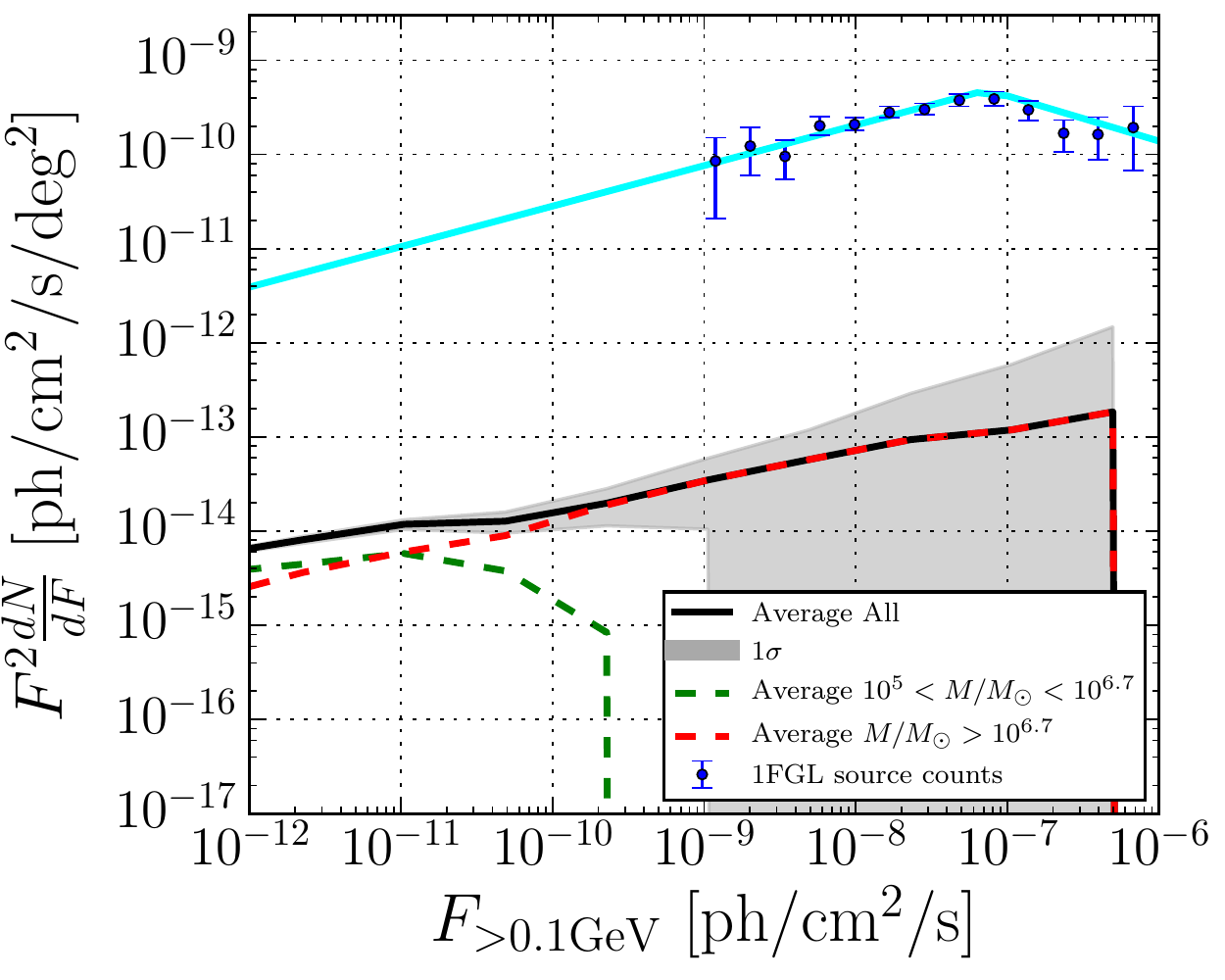}
\caption{Same as left panel in figure~\ref{fig:lognlogsone}, for $M_{\rm SH}\geq10^5 M_{\odot}$ 
	(mass resolution of AQ08). We show separately the contribution of $10^5 M_{\odot} \leq
	M_{\rm SH} < 10^{6.7} M_{\odot}$ (green dashed line), and $M_{\rm SH}\geq10^{6.7} 
	M_{\odot}$ (mass resolution of Hydro AQ, red dashed line).}
\label{fig:logNlogS} 
\end{figure}

Adopting the prescriptions outlined in  section~\ref{sec:DMmodelflux}, 
we simulate on average 38000 SHs from 10$^5$ \Msun~(which is the mass resolution of AQ08) 
up to $5.4 \times 10^6$ \Msun, and
derive $r_s$ from $r_{\max}$ by extrapolating to low masses its polynomial dependence on $M_{\rm SH}$ 
as described in section~\ref{sec:model}.
The results of this new Monte Carlo realization are presented in figure~\ref{fig:logNlogS}, for 
DM annihilating into $b \bar{b}$ and mass $m_{\rm DM} = 100$ GeV,
and for 3FGL catalog setup.
We show the average ${\rm Log\,N}$ -- ${\rm Log\,F}$ of the SHs with masses $\geq 5.4 \times 10^6$ \Msun~
(as resolved by the original simulation)
as red dashed line and the average ${\rm Log\,N}$ -- ${\rm Log\,F}$ of the SHs with masses $10^5 \leq M_{\rm SH} \leq 5.4 \times 10^6$ \Msun~as 
a green dashed line.
The black line shows the total source count distribution from the sum of the two populations of SHs.
For the sake of comparison, we further show as blue line the expected source count distribution 
from blazars in the
1FGL~\cite{Collaboration:2010gqa}. 
Adding lower-mass SHs increases the number of sources per unit flux at very small fluxes. This fact 
has no  impact on the number of 
detectable SHs nor on the constraints on the annihilation cross section.
This result is consistent with figure~\ref{fig:dmasssub}, where SHs with 
$M_{\rm SH} < 10^7$ \Msun~(thus well above the mass resolution of the simulation) are not detectable
as point sources.
Nonetheless, those SHs would unavoidably contribute to the diffuse gamma-ray emission
~\cite{2008Natur.456...73S}. Although very challenging because of the many theoretical uncertainties, and of the unavoidable contribution from unresolved blazars and Misaligned AGN (see e.g. \cite{DiMauro:2013zfa,DiMauro:2013xta}), it is possible to look for those unresolved SHs in the intensity~\cite{Ackermann:2014usa} and small scale 
fluctuations~\cite{Fornasa:2016ohl} of the gamma-ray sky. 
While we do not address this search here, it will be certainly an
interesting topic to explore in future work.

\section{Spatial extension of dark matter sub-halos}
\label{sec:extension}
In this section we discuss one of the clearest signatures for the detection of a DM SH as 
a gamma-ray source: its spatial extension.
Indeed, should an unassociated source be detected by the LAT with a non-zero spatial 
extension at high latitude, it would be a tantalizing hint of a signal from DM SH. Up to 
now only associated astrophysical objects have been 
 found as extended,  and no unassociated object has been detected with a spatial extension.
Estimations of the number of extended SHs that could be detected in the 3FGL have been 
performed in previous works comparing the scale radius $r_s$ with the size of the PSF. 
Ref.~\cite{Schoonenberg:2016aml}, for example, employs the parameter $R_{\rm{ang}} = \arctan{(r_s/d_{\rm{SH}})}$ 
to perform the analysis of the spatial extension. This parameter represents the angular 
size associated to the scale radius of the SH. $R_{\rm{ang}}$ is then compared to the size of 
the PSF for {\tt P7REP\_SOURCE\_V15}, which at 1 GeV is 
$0.8^{\circ}$.
Nevertheless, the definition of $R_{\rm{ang}}$ is not precisely comparable with the way 
extended sources are studied in the 3FGL catalog.
Indeed the size of 3FGL extended sources is determined as the angle
$\Theta_{\rm{ext}}$ inside which the 68$\%$ of the gamma-ray intensity is contained.  

We calculate here the gamma-ray flux for different angular distances from the center of each DM SH and we infer the angular distance $d^{68}_{\rm{SH}}$ 
inside which the $68\%$ of the gamma-ray flux is contained.
We choose a different approach also to estimate the sensitivity of the LAT to detect extended sources.
We use the extension of 3FGL sources and the error on their position to estimate the angular extension sensitivity of the telescope.
First of all, we note that the angular extension of the least extended 3FGL sources is between 
$0.14^{\circ} - 0.20^{\circ}$ for W44 and $0.16^{\circ}$  for HESS J1303-631~\cite{Acero:2015gva}. 
We can then infer the error on the determination of the position of 3FGL sources (at $|b|>20^{\circ}$), 
using the parameter {\tt Conf\_68\_Semiminor},
 reported in {\Fermi}-LAT catalogs (see e.g. \cite{Acero:2015gva}) to parametrize the $68\%$ confidence 
 level of the dimension of the source if modeled with an ellipse.
This parameter is $\sim 0.10^{\circ}$ for most sources with TS = 25 and $|b|>20^{\circ}$. This 
value can be used as an estimation of the lower limit on the spatial extension of a source that can be found in the 3FGL.

We follow two approaches: A conservative one, where we take as a reference angle for the 
SH spatial extension the size of W44 ($\Theta_{\rm{ext}}=0.16^{\circ}$), and a more optimistic choice 
where we consider the average value of {\tt Conf\_68\_Semiminor} for sources with TS = 25 in the 3FGL ($\Theta_{\rm{ext}}=0.10^{\circ}$).
The latter choice is nevertheless not too optimistic, if we consider that with Pass 8 PSF Type 3 (the PSF 
quality quartile with the best angular resolution\footnote{\url{https://www.slac.stanford.edu/exp/glast/groups/canda/lat_Performance.htm}}) 
there is an improvement with respect to the 3FGL (Pass 7) of at least a factor of two in angular resolution. 
If $d^{68}_{\rm{SH}}$ is larger than $\Theta_{\rm{ext}}$, then the SH is considered extended.
We analyze only SHs with a flux larger than the sensitivity flux threshold derived for the 3FGL catalog setup.

Working with all 100 Monte Carlo realizations, a DM mass $M_{\rm DM} = 40$ GeV, annihilation into $b \bar{b}$ and the thermal cross section we have on average, for each realization, 0.5 extended sources when conditioned to $\Theta_{\rm{ext}}=0.16^{\circ}$, while using the optimistic approach 
($\Theta_{\rm{ext}}=0.10^{\circ}$) we get 0.8 extended sources per realization.
These estimated numbers for extended sources in the 3FGL catalog are smaller than 
what has been derived in ref.~\cite{Schoonenberg:2016aml}, where 4 extended objects were predicted 
(assuming $M_{\rm{DM}}=40$ GeV and thermal annihilation cross section).
Indeed, the sensitivity flux threshold used in the analysis performed by~\cite{Schoonenberg:2016aml} 
is different and, as we have shown in the previous sections, this brings to different predictions in the number of detectable SHs.
We are as well using a different approach to define whether a DM SH can be detected as extended source. 

The DM SHs that we find to be extended show the following features: $M_{\rm{SH}}>2\cdot10^{7}M_{\odot}$ 
and distance $<80$ kpc. On average, the smaller is the mass of the 
extended SH, the smaller is the distance. For example, SHs with $M_{\rm{SH}}\sim 1\cdot10^{8}M_{\odot}$ 
are at most located on average at 30 kpc of distance, while less massive objects with $M_{\rm{SH}}\sim 2\cdot10^{7}M_{\odot}$ can be as far as 15 kpc.

\begin{figure}
	\centering
	\includegraphics[width=0.49\columnwidth]{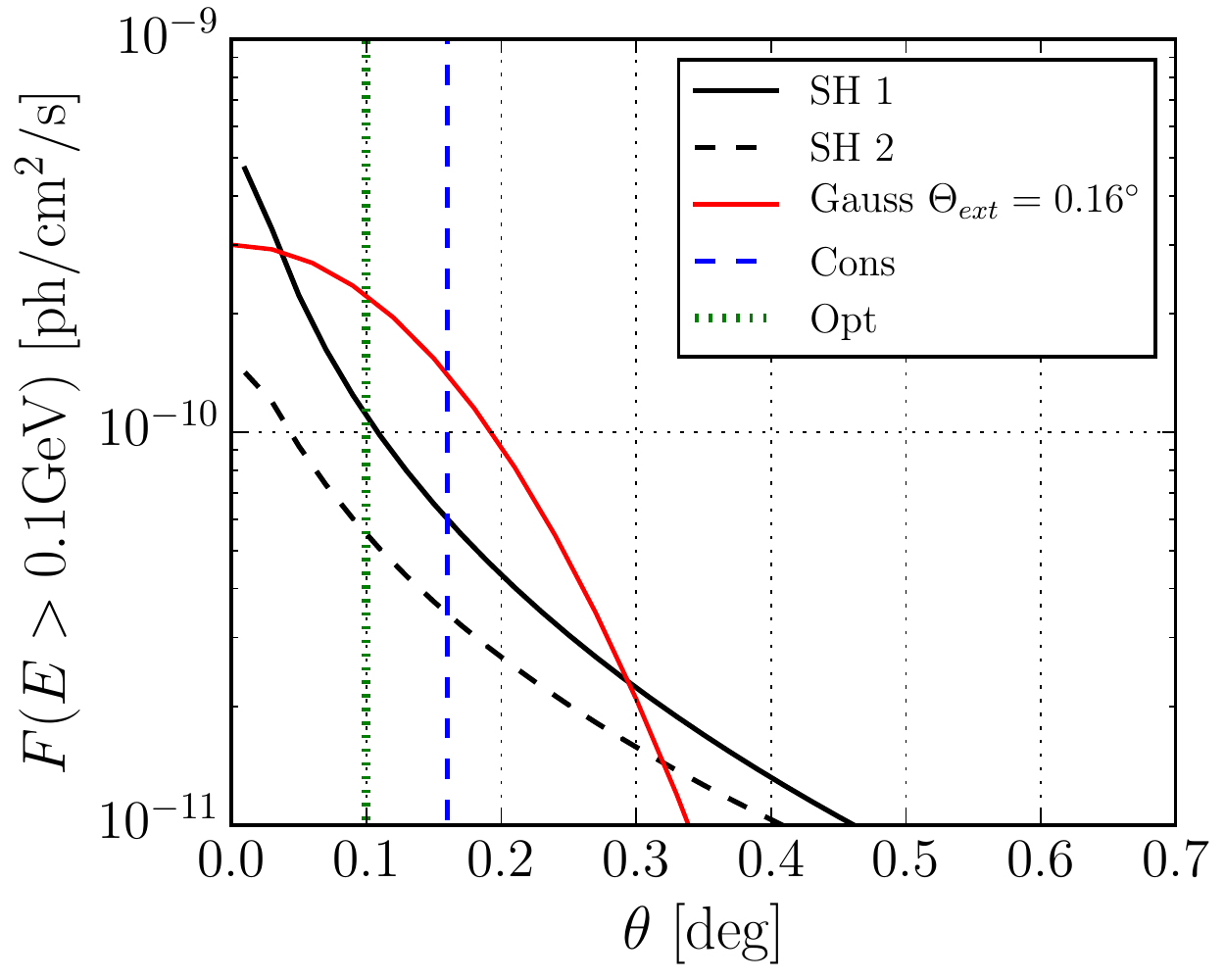}
\caption{Flux profile as a function of the angular separation from the center of two extended DM SHs with 
	a flux larger than the sensitivity flux threshold of the 3FGL setup: $M_{\rm{SH}}=1.9\cdot 10^{9}M_{\odot}$, $r_s = 1.1$ 
	kpc and $d_{\rm SH}$ =46 kpc (solid black line), and $M_{\rm{SH}}=4.7\cdot 10^{9}M_{\odot}$, $r_s = 1.4$ 
	kpc and $d_{\rm SH}$ = 80 kpc (dashed black line).  
	The red solid line corresponds to a Gaussian profile with  $\Theta_{\rm{ext}} = 0.16^{\circ}$.	
	The optimistic and conservative $\Theta_{\rm{ext}}$ are highlighted by the vertical green and blue line respectively.
}
\label{fig:ext} 
\end{figure}

In figure~\ref{fig:ext} we show the flux profile as a function of the angular separation for two 
extended SHs: the first (SH 1) has a mass $M_{\rm{SH}}=1.9\cdot 10^{9}M_{\odot}$, $r_s = 1.1$ 
kpc and $d_{\rm SH}$ =46 kpc, the second (SH 2) has a mass $M_{\rm{SH}}=4.7\cdot 10^{9}M_{\odot}$, $r_s = 1.4$ kpc and $d_{\rm SH}$ = 80 kpc.
We highlight in the same plot the angular distance $\Theta_{\rm{ext}}$ for our optimistic and conservative scenarios.
The angular profile for DM SHs has a steeply decreasing shape that is much different than a Gaussian profile, as it can be 
seen from the figure.  This intrinsic distribution, once convolved with the LAT PSF, would show the 
 sharp peak smoothed over a larger solid angle,  making the angular emission more similar to a Gaussian function.
The presence (and shape) of the extension by itself is not sufficient to claim an evidence of DM SH. 
Would a source be detected as extended, it should be an unassociated source 
in the 3FGL \emph{and} future \Fermi-LAT catalogs before being claimed a possible DM SH . 
An additional remark is that given the improvement in the LAT sensitivity with Pass 8, future catalogs 
will contain many more sources than the 3FGL catalog. With an increasing number of detected sources, 
the probability of having two gamma-ray sources detected with a distance of the order of the LAT PSF, 
and thus looking as a single extended source,  
is not negligible. 
This hypothesis therefore should be considered when an extended unassociated \Fermi-LAT source 
will be discovered, and even more if the source spectrum will show a good match with a DM-like spectrum.

\section{Summary and conclusions}
\label{sec:concl}
We have presented a realistic estimation of the detectability of Galactic dark matter 
sub-halos in the \Fermi-LAT 3FGL and 2FHL catalogs.
Based on one of the most recent hydrodynamic simulations for structure formation, the Hydro-Aquarius 
simulation~\cite{Marinacci:2013mha,Zhu:2015jwa}, 
we have modeled the spatial and mass distribution of sub-halos in a Milky Way-like Galaxy. 
We have generated Monte Carlo realizations of the Galactic sub-halo population (with minimum mass 
$M_{\rm SH} \sim 5 \times 10^{6}$ \Msun) for  the hydrodynamic and pure dark matter scenarios.
Our first motivation was to investigate the impact of hydrodynamics on the distribution and properties 
of Galactic dark matter sub-halos, and consequently on the gamma-ray signal expected from 
those structures. At this scope, we have compared the scale radius typical of each sub-halo, 
deeply related to the sub-halo mass accretion history and to the concentration parameter. 
Being a physical parameter of the radial  sub-halo density, it  is indeed a crucial quantity for 
the determination of the gamma-ray flux. We modeled $r_s$ directly from the simulation data of $r_{\rm max}$.
Although baryons affect the abundance and internal structure of sub-halos (especially the more 
massive ones), these discrepancies do not substantially alter the predictions on  $r_s$. 
This conclusion holds as well for the geometrical factor \Jf, which is a direct measure 
of the intensity of the gamma-ray signal. 
 
In order to estimate the realistic sensitivity for the \Fermi-LAT  to detect dark matter sub-halos,  
we have introduced some novelties. In particular, we fully account for dependence of the sensitivity 
flux threshold on the dark matter annihilation channel, the dark matter mass and the sub-halo position 
in the main halo. We have overcome the simplistic approach of considering a fixed sensitivity 
flux threshold, showing in particular the strong dependence of the sensitivity flux threshold on 
the dark matter mass. 
Moreover, we have presented the prospects of detection of sub-halos among the unassociated 
sources of two \Fermi-LAT catalogs: probing different energy ranges, the results for 3FGL and 
2FHL result complementary.

We have studied the dark matter annihilation gamma-ray signatures, from Galactic sub-halos in terms of: 
(1) the number of detectable sub-halos in the two catalogs, (2) the bounds on the dark matter 
annihilation cross section, (3) the source count distribution and (4) the sub-halos extension.
Our results show that the largest number of detectable sub-halos, that might already be among 
the unassociated sources of the 3FGL catalog, is at most $0.9\pm0.8$ for $M_{\rm{DM}}=8$ GeV 
-- with $\sv$ fixed to the upper limit derived from the latest analysis of dwarf spheroidal galaxies. 
The prediction for the 2FHL catalog is lower: $N_{\rm{Detectable}} = 0.0\pm0.2$ for $M_{\rm{DM}}=10$ TeV. 
These tiny numbers allow to set constraints on the dark matter annihilation cross section into gamma rays. 
Although the upper limits on $\sv$ for the detection of $N_{\rm Candidate}$ = 5 or 20 sub-halos are 
weaker than those derived with the Pass 8 analysis of dwarf galaxies~\cite{Ackermann:2015zua}, they become quite competitive assuming zero sub-halo candidates.
For values of $\sv$ consistent with the current limits from dwarf galaxies, we have also 
found that the sub-halos source count distribution is  suppressed 
by more than three orders of magnitude with respect to the observed flux distribution of 
blazars in both the 3FGL and 2FHL catalogs. Moreover, we have investigated the impact of 
adding smaller mass sub-halos ($10^5$ \Msun $< M_{\rm SH}< 5 \times 10^{6}$ \Msun) to the 
sub-halo population. 
Their effect is to  increase the number of sources per unit flux at very small fluxes. As a 
consequence, they have no effect
on the number of detectable sub-halos and on the bounds to the annihilation cross section 
for current sensitivities.

 One discriminating feature for the identification of dark matter sub-halos would be the spatial 
 extension of the source. About one sub-halo
 of our simulated population turns out to be detectable in the 3FGL as extended source.
We recall that conservative assumptions have been made in the present work. Indeed, we
 expect  a great improvement  with the new Pass 8 4FGL, which could significantly increase 
 the number of detectable sub-halos, and possibly lead to the identification of 
some unassociated sources with dark matter halo substructures thanks to their spatial extent. 
As an illustrative example of future progresses, we have considered a 
 gamma-ray instrument with a factor of 5 better sensitivity than the LAT above 100 MeV, 
an improvement that can be achieved by new concept MeV telescopes such as
e-ASTROGAM~\cite{DeAngelis:2016slk} and Compair~\cite{Moiseev:2015lva}.
Given the input of the adopted simulation, we have found that it will be possible 
to detect a few sub-halos (about 2.1), with a probability of 75$\%$ (1.5/2.1) to detect a dwarf galaxy.

We finally also note the relentless efforts in the numerical simulations for the reliable inclusion 
of the effects of baryons in the formation of galactic structures. These latter research leaves room 
open to further inspections, once Milky Way size halos will be realized with even greater resolution. 

\section*{Acknowledgments}
We acknowledge fruitful discussions with G.~Bertone, D.~Schoonenberg, Q.~Zhu, H.~S.~Zechlin, S.~Palomares-Ruiz, 
E.~Charles and M.~Wood. We also thank V.~Springel for comments on the manuscript.
FC acknowledges partial support (for the initial stages of this work) from the European Research Council 
through the ERC starting grant WIMPs Kairos, P.I. G.~Bertone.
VDR acknowledges support by the Spanish MINECO through the project FPA2012-31880 
(P.I. Enrique Alvarez Vazquez) and partial support from the EU's Horizon 2020 research and 
innovation programme under the Marie Sklodowska-Curie grant agreements No 690575 and No 674896.  
MDM acknowledges support by the NASA Fermi Guest Investigator Program 2014 through the Fermi 
multi-year Large Program N. 81303 (P.I. E.~Charles). 
FC and FM would like to express a special thank to the Mainz Institute for Theoretical 
Physics (MITP) for its hospitality and support.

\clearpage
\bibliographystyle{JHEP}
\bibliography{draft}

\end{document}